\documentclass[10pt]{article}
\pdfoutput=1
\usepackage{graphicx} % Required for the inclusion of images
\usepackage{amsmath} % Required for some math elements 
\usepackage{amsfonts}
\usepackage{amssymb}
\usepackage{subfigure}
\usepackage{multirow}

\title{ A new Phenomenon: Glass Paramagnetism. \\
Further Experimental and Theoretical Details  } % Title

\author{ Giancarlo Jug$^{a,b,c}$ and Sandro Recchia$^a$ \\ \\
$^a$Dipartimento di Scienza ed Alta Tecnologia \\ 
Universit\`a dell'Insubria, Via Valleggio 11, 22100 Como (Italy) \\
$^b$INFN -- Sezione di Pavia, Italy \\
$^c$ To.Sca.Lab, Dipartimento di Scienza ed Alta Tecnologia, Universit\`a dell'Insubria \\
}

\date{\today} % Date for the report

\begin{document}

\maketitle

\begin{abstract}
% Abstract text
{\bf
In a recent manuscript, the discovery of a new phenomenon in glasses has been 
reported: {\it glass paramagnetism}, that is the intrinsic magnetisation developed 
by a substance that has failed to crystallise in a temperature quench when placed 
in an external magnetic field. The field- ($H$) and especially the temperature- 
($T$) dependence of the intrinsic magnetisation $M_{intr}=M_{intr}(H,T)$ is 
very unexpected, with broad peaks developing for high $H/T$ and marked $T$-
oscillations at fixed $H$ for intermediate-to-high temperatures. 
In this work we present details on the samples, new data for other glassy systems 
and especially the theoretical background that is capable of explaining most of the 
experimental data. New phenomena are however emerging from these data, for 
example a deviation from the Curie-like behaviour of the magnetic susceptibility 
($M_{intr}/H$ at low fields) due to the glassy structure alone as well as evidence 
for new and intriguing quantum coherence effects at the lowest temperatures.
}
% Abstract text
\end{abstract}

\tableofcontents
A. The Samples \\
B. Polycluster and Extended Tunneling Models for Glasses \\
C. Some More Data Fitting Results \\
D. Further Discussion about Curie Law and Comparison with ETM-Theory \\
E. The $M_{intr}$ Temperature Oscillations and the High Field Limit \\
F. Novel Quantum-Coherence Phenomena \\
G. Final Conclusions, Authors and Acknowledgements

\vskip 1.0cm
%\begin{center}
%{\em ''Beware ye, all those of bold spirit who want to suggest new ideas!'' (B.D. Josephson)}
%\end{center}

%\begin{abstract}
%% Abstract text
%\noindent
%{\bf
%Supplementary information to main article. Samples, theory model details, 
%further data analysis and further discussion of results and comparison with theory.
%}
%% Abstract text
%\end{abstract}

%\vfill
%\end{document}

\newpage

%----------------------------------------------------------------------------------------
%	SECTION 1
%----------------------------------------------------------------------------------------

\begin{figure}[!h]
\centering
{ \vskip -0cm
   \subfigure[]{\includegraphics[scale=0.045]  {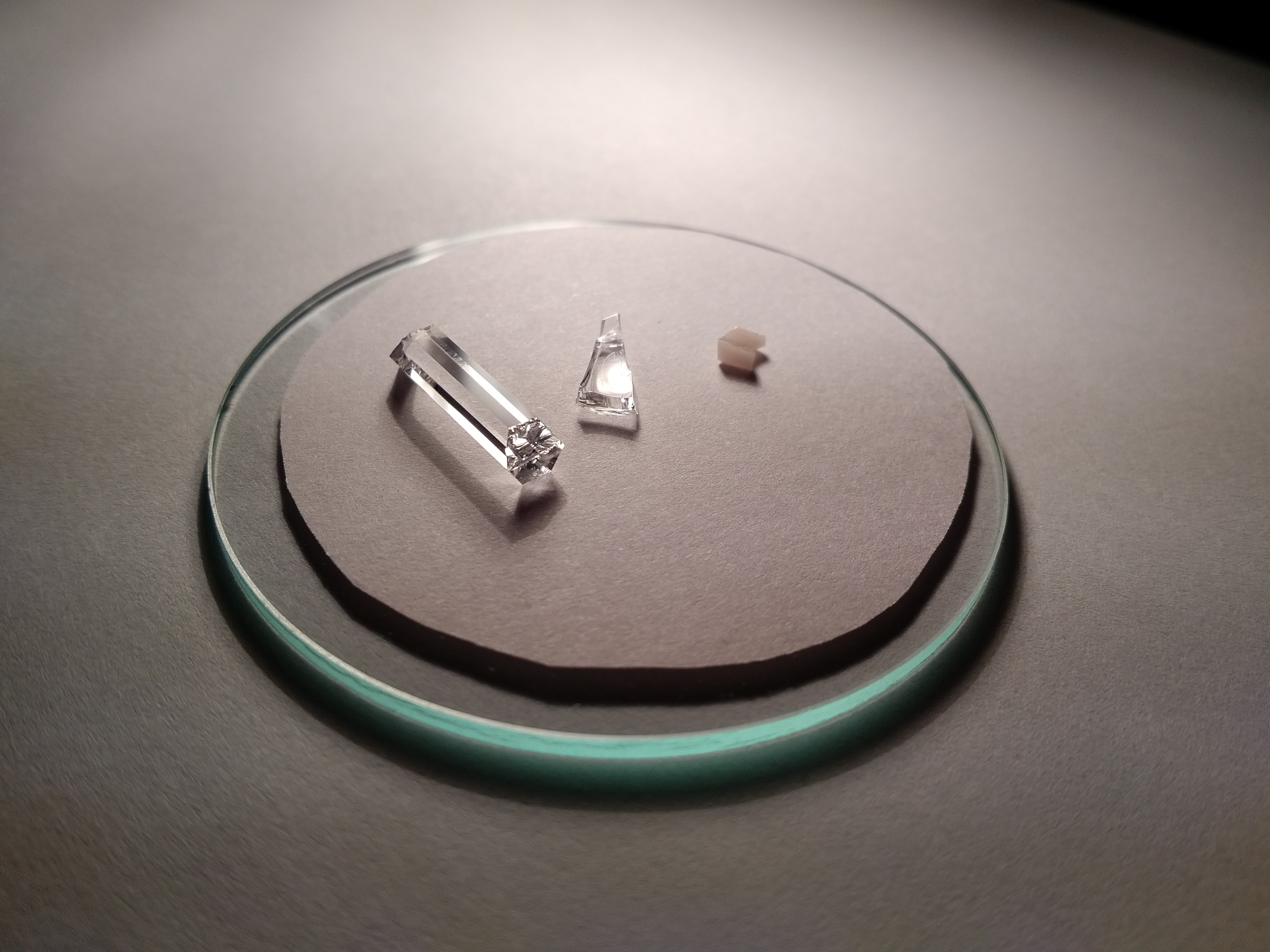} } 
 \vskip -0cm
   \subfigure[]{\includegraphics[scale=0.03] {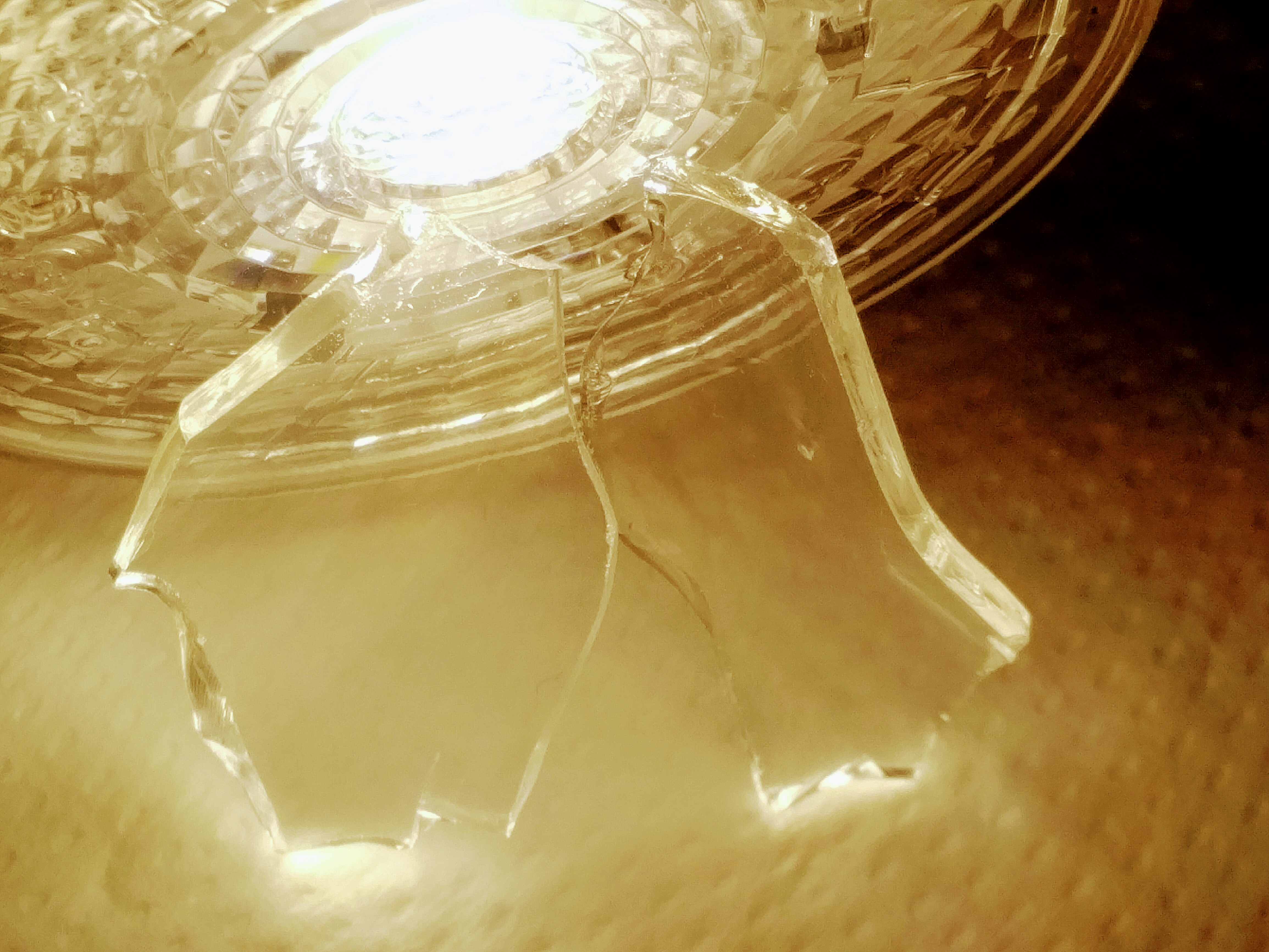} }
 \vskip -0cm
   \subfigure[]{\includegraphics[scale=0.03] {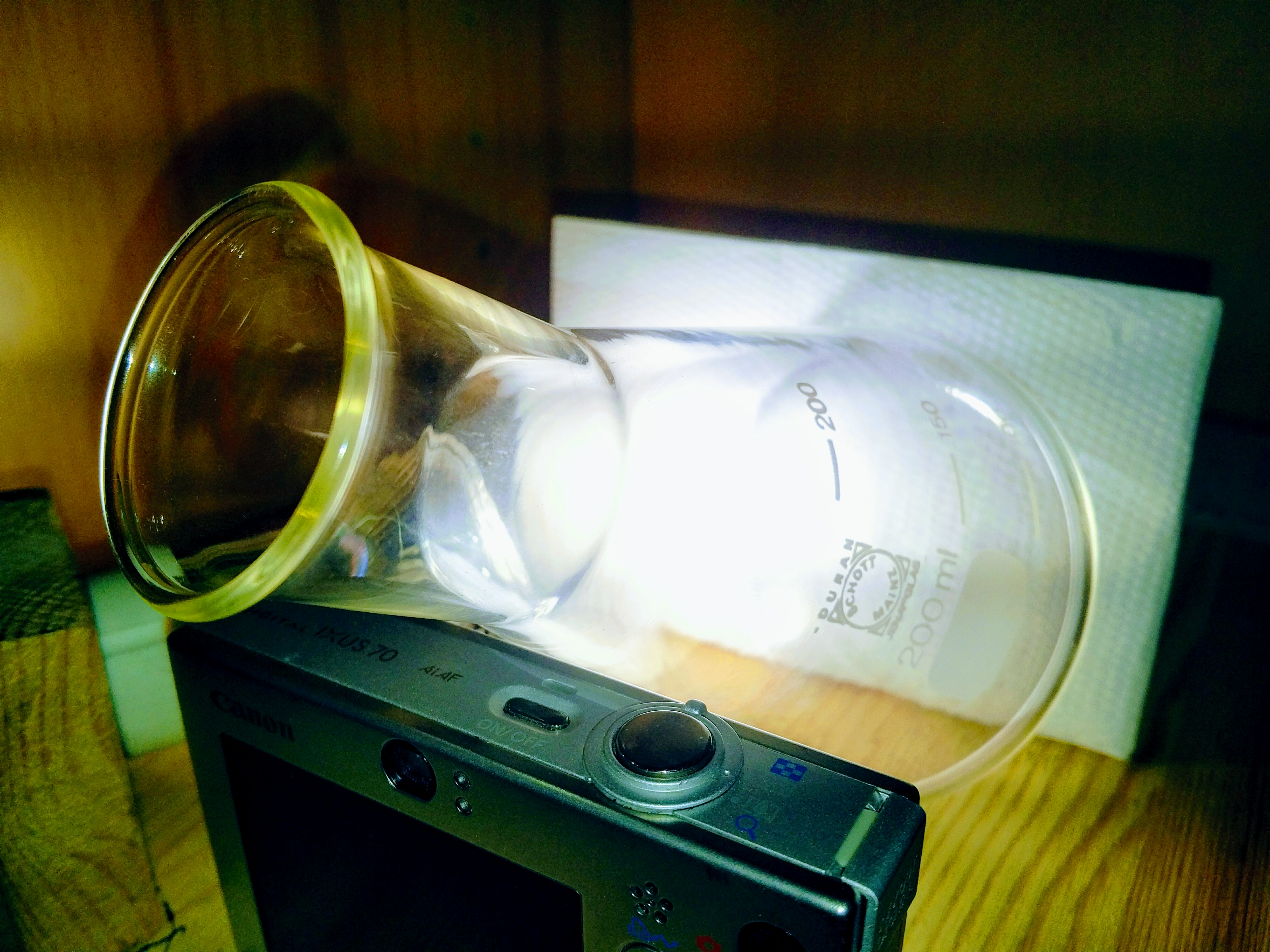} }
}
\caption{ 
(a) Some of our glass samples (from the left: BK7 (remaining prism), Duran 79.5 mg, 
BAS-p  27.9 mg, as explained in the PA text). The base-lens (made of ordinary window 
glass, diameter 57 mm) is displayed to show the green coloration due to 
Fe$^{2+}$-predominance. 
(b) Shards from the very same Duran glass beaker from which we obtained the sample
in the middle of (a); the greenish tinge of this glass is clearly visible along the edges. 
(c) Schott's Duran beaker, an identical copy of which we shattered to make our Duran
samples; noticeable is the green coloration along the longest optical path (also visible 
in the Duran shard at the centre of (a), front upper rhs-corner and front bottom edge,
and along the edges of the shards in (b)). Illumination was with white LED light.
}
\label{samples}
\end{figure}

{\bf Introduction.} This manuscript is the continuation of a shorter, parent article and 
readers are referred to the Introduction of that article for background information.
Hereafter, the Parent Article \cite{JR2021} is throughout indicated as the PA; MS stands 
for Mass Spectroscopy.  PI is the Principal Investigator (GJ).
\vfill
{\bf A. The Samples.} Fig. \ref{samples}(a) shows three of our four glass samples, 
measured in the Quantum Design SQUID magnetometer. Together with chips from 
the BK7 prism (what remains of it) on the left, we produced shards from a 
shattered Schott's Duran beaker identical to the one pictured in Fig. \ref{samples}(c). 
We insist on this issue, the green coloration, because it invalidates claims made
from naive LL SQUID-measurements that Fe is contained only as Fe$^{3+}$ (coloring 
yellow) in ordinary multi-silicate glasses. It is not, it is predominantly Fe$^{2+}$ (coloring
green) and this enhances the discrepancy between the LL-SQUID-ascertained $n$(Fe) 
and the true $n_{MS}$(Fe) value. This strenghtens the case for the 
existence of an intrinsic, glassy state magnetisation.

%----------------------------------------------------------------------------------------
%	SECTION 2 
%----------------------------------------------------------------------------------------

{\bf B. Polycluster and  Extended Tunneling Models for Glasses. } Alternatives to the 
popular Zachariasen-Warren CRN model of glasses \cite{Zac1932,War1934} have 
been occasionally proposed in the literature. The cybotactic-groupings model of
Valenkov-Porai-Koshits-Lebedev \cite{VPK1936,Leb1940} (for recent reviews, 
see \cite{Por1990,Gas1998,Wri2014}), the modified-CRN model of Greaves 
\cite{Gre1985} and the polycluster model of Bakai \cite{Bak1994,Bak2013} are the 
closest in spirit to the one independently proposed by the PI \cite{Jug2018}, 
except that the cybotactic and Greaves versions both consider better-ordered and 
less-ordered regions as highly inter-penetrating each other. 
In the polycluster model, like in the PI's own \cite{Jug2018}, the better-ordered 
regions are solid-like compact objects that get to be jammed together below 
$T_g$ and contain ``voids'' filled with mobile, fluid-like particles (ions or molecules) 
from the surviving supercooled melt. In a nutshell, the clusters or RER are failed 
nanocrystals almost monodispersed in size and randomly closed-packed at low 
enough $T$, their network percolating the entire system (see Fig. 5(a) in the PA and 
Fig. \ref{hrtem} below). Vogel in Refs. \cite{Vog1982,Vog1992} and especially in 
Ref. \cite{Vog1971} (see also Section D)  as well as Zarzycky \cite{Zar1991} 
have provided dramatically convincing HRTEM images of the hidden granularity of 
glasses. See however \cite{JLT2021} for a distinction between truly dynamical (ca. 1 
nm size) heterogeneities -- characterising single-component systems -- and quasi-static
heterogeneities (10s to 100s of nm in size) typical of the multi-component systems.

We focus on the fluid-like particles from the remaining supercooled melt 
in the ``voids'', which are of a highly complex nature, especially in the 
multi-component glasses where micro-phase separation between 
network-forming (NF, e.g. SiO$_2$, Al$_2$O$_3$, B$_2$O$_3$, ...) and 
network-modifying (NM, e.g. Na$_2$O, K$_2$O, BaO, ...) components has long 
been known to take place \cite{Vog1982,Vog1992,Zar1991}. A couple more 
HRTEM pictures are in Fig. \ref{hrtem}.
The idea, then, is that the NM (good crystal-formers') atomic species tend to 
segregate in the RER while the NF (good glass-formers') species tend to segregate 
-- locally and only in part -- in the voids. In any event (even if the opposite happens), 
it is natural to believe that a highly-heterogeneous schematic situation like that in 
Fig. 6(b) in the PA -- when needed -- better characterises glass 
structure at medium-range (or mesoscopic) scales than the CNR model. We stress that
the polycluster model is clearly only an effective model, glossing over considerable (and 
unexplored) detail.

\begin{figure}[!h]
\centering
{ \vskip -0.5cm
 \subfigure[]{\includegraphics[scale=0.6]  {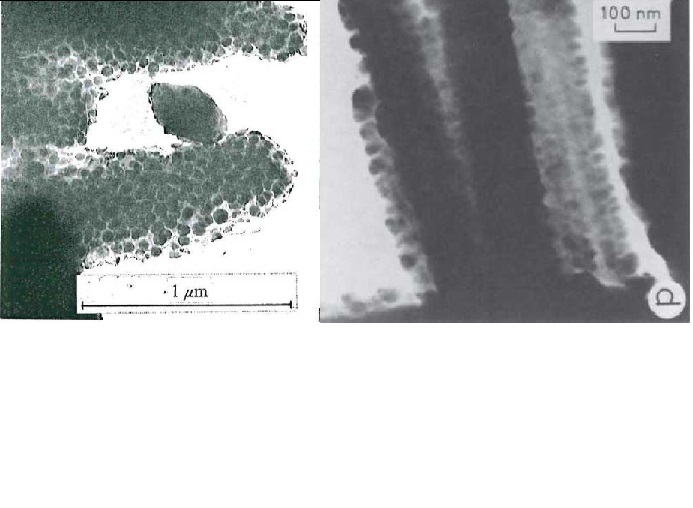} }
\vskip +1.5cm
 \subfigure[]{\includegraphics[scale=0.5] {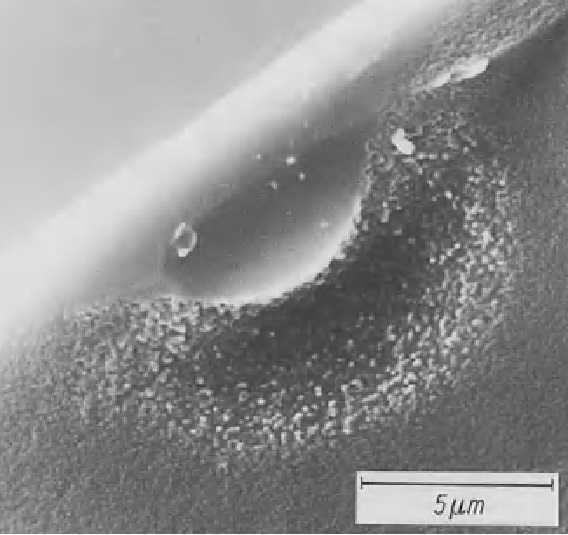} }
\vskip+2.5cm
}
\vskip -2cm
\caption{ (a) Two more HRTEM images showing the polycluster structure of network 
glasses. Left panel: the (B$_2$O$_3$)$_{0.75}$(PbO)$_{0.25}$ (wt\%) glass 
system, its granularity being exacerbated by micro-phase separation. Right panel: 
a-SiO$_2$. In both cases sample preparation in the form of thin slices for 
EM-imaging enhances, through partial stress-release, the size of clusters with respect 
to bulk systems. From \cite{Zar1991}. (b) The HRETM image of the fractured surface 
of the same Li$_2$O-SiO$_2$ glass as in Fig. 6(a) of the PA: the granularity of the 
glass' bulk as well as the cellular structure on the fracture's surface are quite evident.
From \cite{Vog1992}.
}
\label{hrtem}
\end{figure}

Next, the shape and size of each void, or pore, between four RER (tetrahedral void) 
can be estimated from HRTEM images like in Fig. 6(a) (PA)  and in 
Fig. \ref{hrtem} -- or, better, from fitting low-temperature heat-capacity and 
magnetisation data to this theory -- to be of distorted tetrahedral minimal shape and 
having volume (if $2\xi$ is the size of a typical RER):
\begin{equation}
V_{t-void}=\big\{ \frac{2\sqrt{2}}{3}-\frac{4}{3}\arccos{\big( \frac{23}{27} \big)} 
\big\}~\xi^3 \simeq 0.208~\xi^3.
\end{equation} 
%[verificare!]
This for $\xi\simeq$ 30 nm, a typical value for BK7 as it turns out \cite{Jug2018}, 
gives a volume of some 5.6 $\times$ 10$^6$ $\AA^3$ and hence a linear size of 
some 3.62 $\times$ 10$^2$ $\AA$ for the tetrahedron's side. This is enough to 
accommodate some 400 atomic species on each tetrahedron void's face in a packed 
monolayer fashion. These species can be reasonably considered to be O$^-$ species 
adsorbed on each face from the melt in the void, and these negative particles will be 
highly correlated and electrostatically interacting, so that not all available tetrahedron-
void face's adsorption sites will be occupied. Let us say that only some $N_{tunn}$ will 
be present as dangling O$^-$ bonds on each face. Values of $N_{tunn}$ in the range 
20 to 600 have in fact been extracted from applications of the Extended Tunneling 
Model (ETM) below to a plethora of low-temperature experimental data 
\cite{Jug2018,Jug2013,JBK2016}. 
The void's shape is clearly only topologically tetrahedral, distortion is everywhere.

\begin{figure}[!h]
\centering
{ \vskip -1cm
   \subfigure[]{\includegraphics[scale=0.4]  {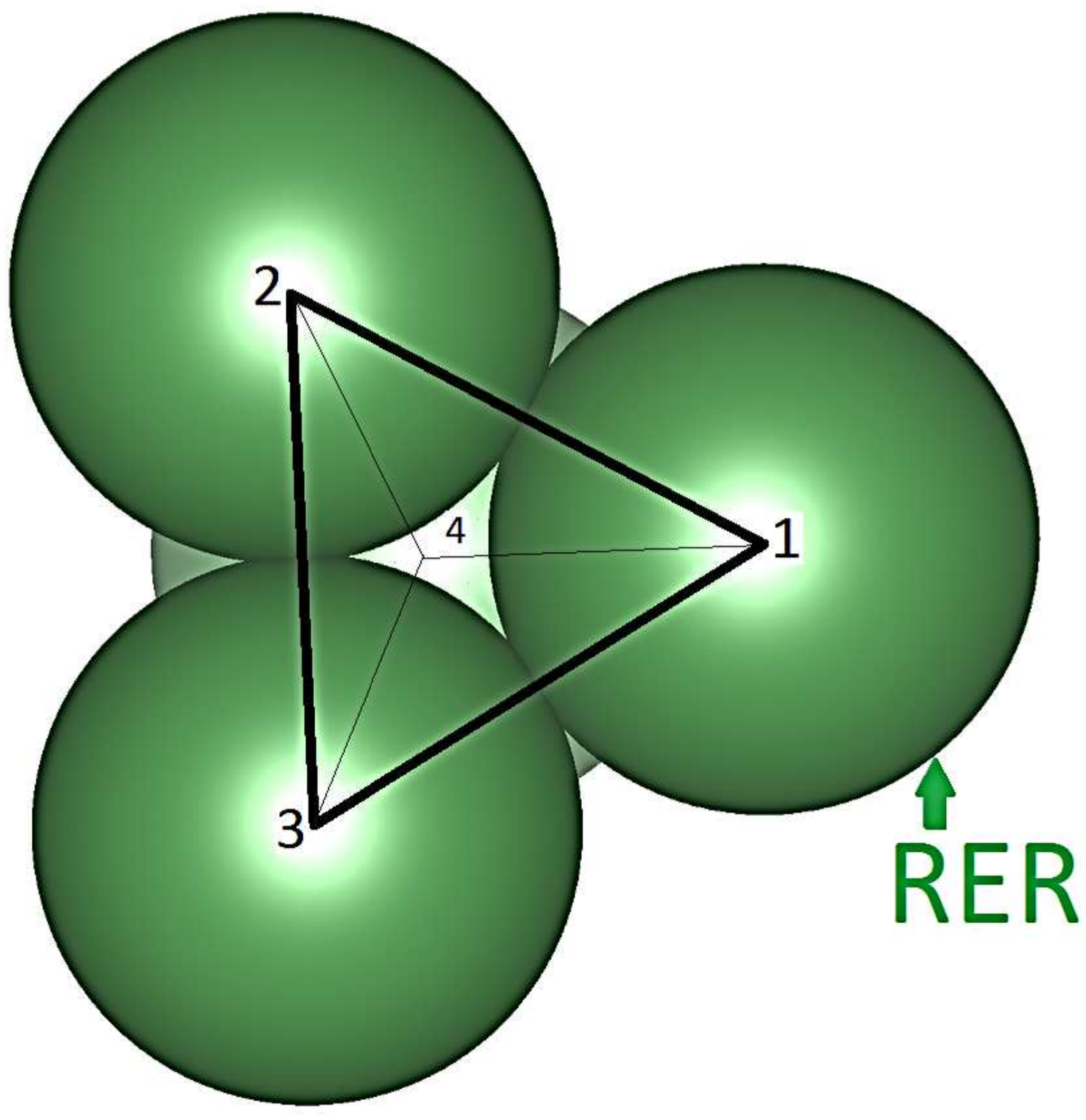} } 
 \vskip -0.5cm
   \subfigure[]{\includegraphics[scale=0.50] {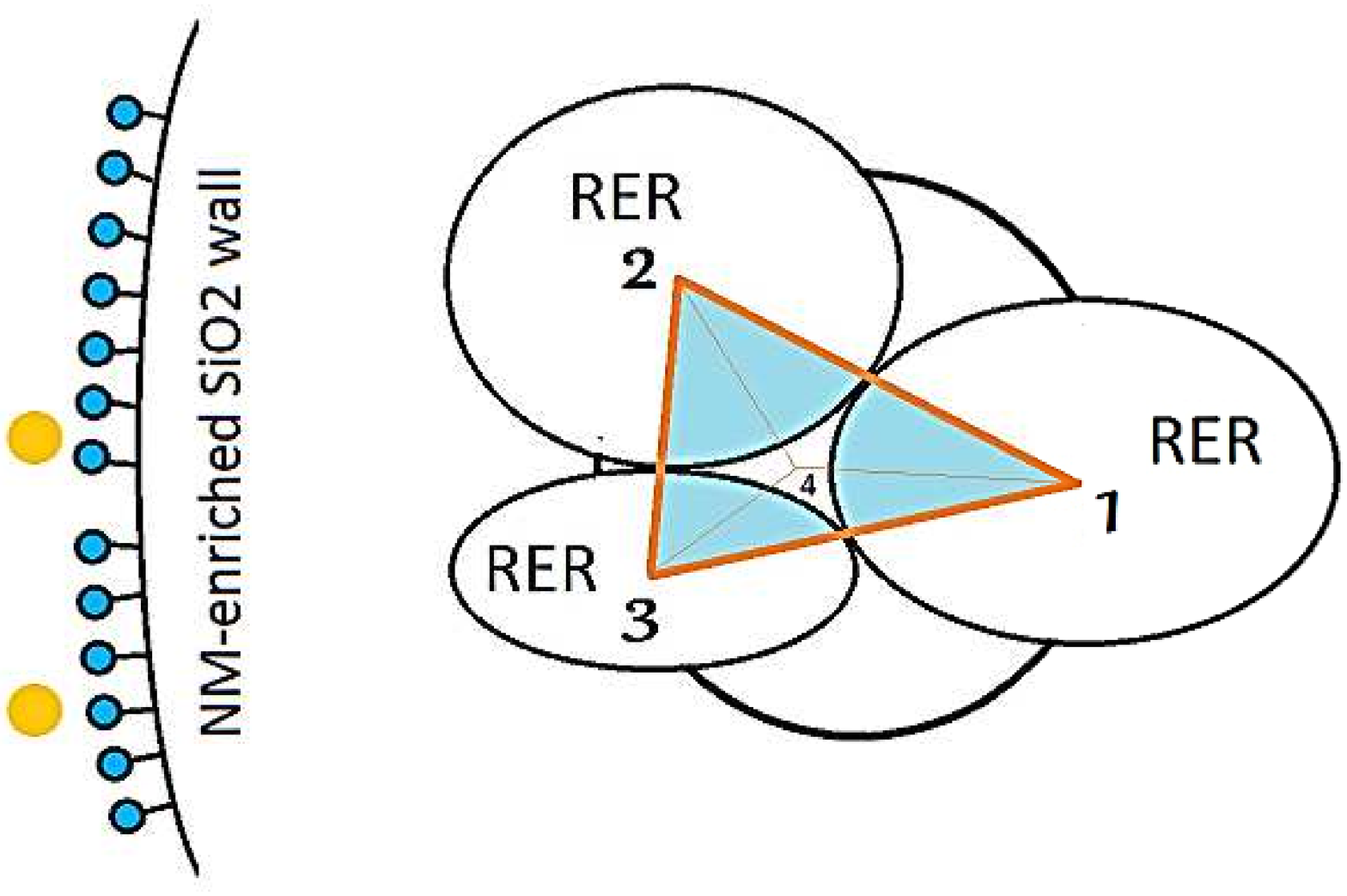} }
}
\vskip -0cm
\caption{ 
(a) tetrahedral void (octahedric and cubic voids are also present, but ignored for now)
between RER, idealised as perfect spheres (which they are not). The actual space 
available to the mobile particles is that of the tetrahedron having vertices 1-2-3-4 
minus the space occupied by the RER inside the tetrahedron. This space has volume
$V_{t-void}\simeq 0.208~\xi^3$ where $\xi$ is the typical ideal-RER ``radius'' 
($2\xi$ is its linear size). (b) Schematic representation of RER void's coating by 
O$^-$ species (dangling bonds on light blue surfaces of more realistic, ellipsoidal RER 
shapes, delimiting the void) as exemplified in the particular drawing on the left: blue 
species O$^-$, golden species Si$^{4+}$ or other NM-cation approaching and 
eventually adsorbing on the RER wall, causing it to grow and to attract replacement 
O$^-$ from the surviving melt in the void. There is naturally a coating (not shown) on
RER n.4 too.
}
\label{ATS-nature}
\end{figure}

\begin{figure}[!h]
\centering
{ \vskip -0cm
\includegraphics[scale=0.6]  {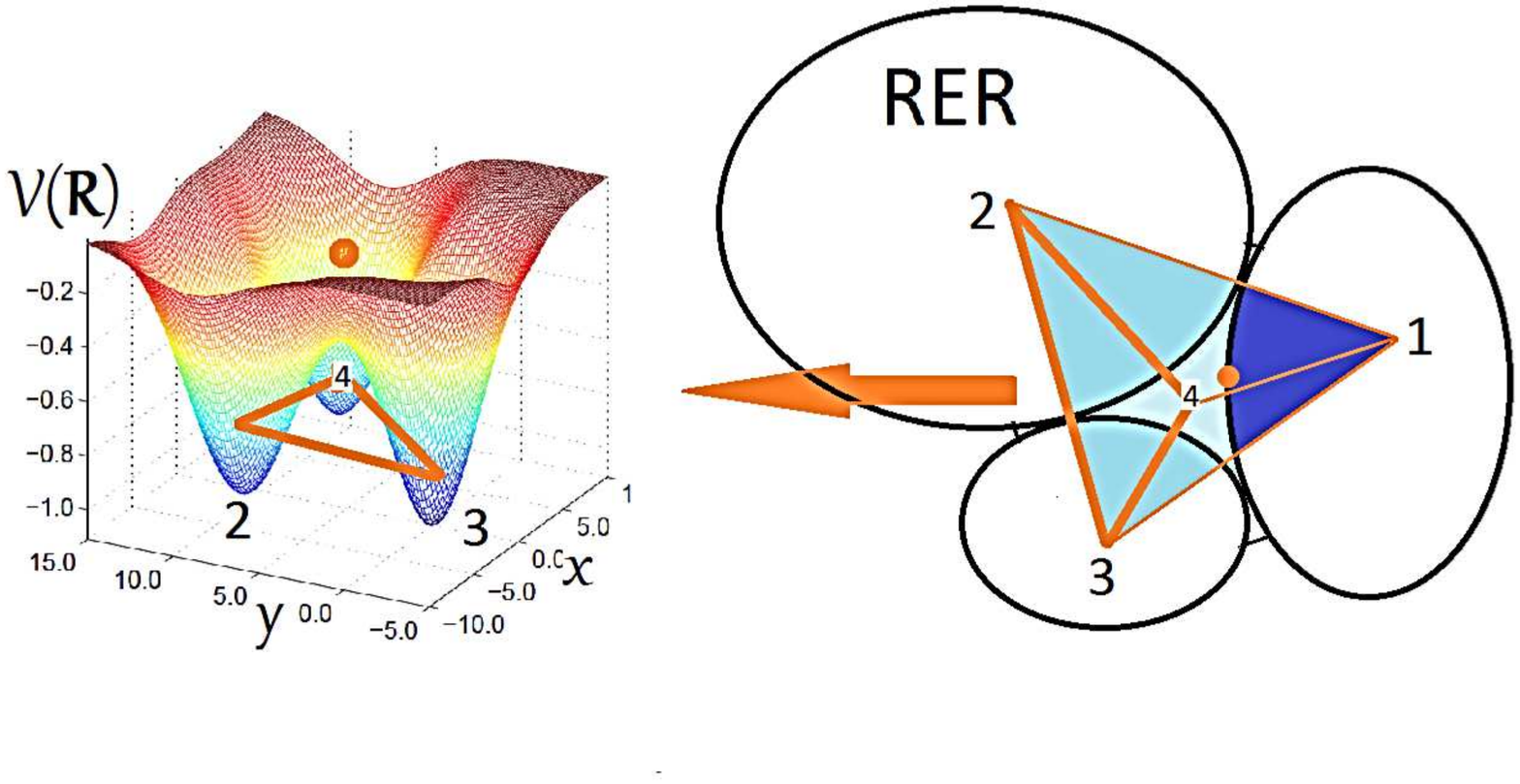}  
}
\vskip -5cm
\caption{ 
Origin of the tetrahedral void's effective triple-welled potential. Right panel: charged ions
(probably O$^-$ dangling bonds) adsorb on the RER surfaces delimiting a given void.
One can consider the large ensemble of ions on the relevant surface of one RER (say 
n.1,  deep blue) as replaced by a quasi-particle (orange dot) interacting with the complex 
distribution of all other ions (in the melt inside the void and on the other three RER 
surfaces, light blue coverage) in the narrow void's space. As a result, right panel: 
quasi-particle n.1 experiences an effective three-well potential $V{(\bf R})$ from RER 
ns. 2-3-4 (${\bf R}=(x,y)$ is some coordinate in the plane of the 2-3-4-triangle and 
the height $z$ is fixed). The same applies for the effective particles on RER ns. 2, 3 
and 4 which experience, for pure topological reasons, also a three-well potential from 
the opposite triplets of RER. 
}
\label{ATS-potential}
\end{figure}

Fig. \ref{ATS-nature}(a) shows (idealized, in practice all will be random) the packing
of the compact RER (strickly speaking, just below $T_g$ idealisation) and the creation 
of a tetrahedral ``void'' which in practice will be filled with many charged and mobile 
particles. There are octahedric and cubic distorted voids too, but in order to keep 
the theory simple these other void topologies will be ignored for now. The charged 
particles inside each void should be described by the microscopic Hamiltonian:
\begin{equation}
{\cal H}_{micr}=\sum_{i=1}^N \frac{p_i^2}{2m_i}+\sum_{i>j} \frac{q_iq_j}{r_{ij}}
+\sum_{b=1}^4 \frac{q_iq_{J_b}}{R_{i,J_b}}
+\sum_{b,b'} \frac{q_{J_b}q_{J_{b'}}}{R_{J_bJ_{b'}}}+\cdots
\label{micrH}
\end{equation}    
where $i,j$ are atomic particles inside the void, $m_i$ are their masses, $J_b$ labels a 
solid-like particle inside one of the four ``solid clusters'' (the RER) surrounding the 
void, $q_i$, $q_{J_b}$ are the charges of the particles in the void and inside the blobs, 
respectively, and $r_{ij}$, $R_{iJ_b}$, $R_{J_bJ_{b'}}$ are the distances between 
all these pairs of interacting atomic particles. The last term is in practice redundant, as 
the positions of solid-like particles changes little, the third term ought to include all of 
the RER in the system, but the truth of the matter is that such microscopic model is 
clearly intractable (even through computer simulation). Therefore, one resorts to an 
effective Hamiltonian, initially focusing on a group of the particles in the void: those 
deposited on the surface of one RER (e.g. RER n.1 in Fig. \ref{ATS-nature}(b)) which 
are presumably mostly O$^-$ dangling bonds. One then replaces these highly correlated 
particles with a single {\em effective} ``particle'' of mass $M$, charge $Q$ subjected to 
an effective potential $V_{eff}$ that, as shown in Fig. \ref{ATS-potential}, will be 
characterized by three minima in correspondence with the three (n. 2-3-4) RER facing 
the chosen (n.1) effective-particle zone. The impossible Hamiltonian in 
Eq. (\ref{micrH}) then becomes tractable \cite{theory}:
\begin{equation}
{\cal H}_{eff}\approx \frac{P_b^2}{2M}+V_{eff}({\bf R}_b)
\label{effH}
\end{equation}
where ${\bf R}_b=(x,y,z)_b$ is a coordinate for the effective-particle $b$ on the plane 
of the opposite triangular RER-arrangement and $z_b$ its (fixed, in practice) distance 
from that plane. This holds for each ``particle'' $b$ (=1,2,3 or 4) defined on each of
the RER-faces involved in that void. However, it is likely that the character of the 
effective particle will change from surface-like (2D in nature) to void-like (3D) as 
temperature drops considerably (see Section F). Indeed, through adsorption of mobile 
void-species on the RER surfaces, the RER get to grow and compactify at the expense of 
the fluid-like species in the void. Thus, one can expect that at the lowest temperatures 
the effective particle will characterise the whole of the void's atomic particles' distribution 
and the $V_{eff}({\bf R})$ effective potential will acquire a four-welled tetrahedral
configuration. 

At this point one is set to represent the (unknown, except for its topology) effective 
potential $V_{eff}({\bf R})$ in terms of the quantum-mechanical basis 
$|\alpha\rangle$, $\alpha$=1, 2, 3, of its three wells, to get ${\cal H}_{eff}$ in 
matrix form and with a minimal number of parameters:
\begin{equation}
{\cal H}_{3LS}=\langle\alpha | {\cal H}_{eff} | \beta\rangle=
\left( \begin{array}{ccc}
E_1 & D_0 & D_0 \cr
D_0 & E_2 & D_0 \cr
D_0 & D_0 & E_3 \end{array} \right)
\label{3lstunneling}
\end{equation}
the $E_1, E_2, E_3$ being the energy asymmetries between the wells (choosen so 
that $E_1+E_2+E_3=0$ for each tunneling system (TS)) and $D_0$ is the most 
relevant tunneling amplitude (through saddles, in fact). This three-level system 
(3LS) Hamiltonian is  the natural generalisation of the popular Standard Tunneling 
Model's (STM) 2LS Hamiltonian \cite{Esq1998},
\begin{equation}
{\cal H}_{2LS}=\left( \begin{array}{cc}
\Delta/2 & \Delta_0/2 \cr
\Delta_0/2 & -\Delta/2 \end{array} \right)
\label{2lstunneling}
\end{equation}
($\Delta$ being the energy asymmetry and $\Delta_0$/2 the tunneling amplitude) 
and has the advantage of readily allowing for the inclusion of a magnetic field 
$B=\mu_0 H>0$, when coupling orbitally to the tunneling ``particle'' having 
effective charge $Q$: 
\begin{equation}
{\cal H}_{3LS}(B)=\left( \begin{array}{ccc}
E_1 & D_0e^{i\varphi/3} & D_0e^{-i\varphi/3} \cr
D_0e^{-i\varphi/3} & E_2 & D_0e^{i\varphi/3} \cr
D_0e^{i\varphi/3} & D_0e^{-i\varphi/3} & E_3 \end{array} \right).
\label{3lsmagtunneling}
\end{equation}
Here $\varphi/3$ is the Peierls phase for the tunneling particle through a saddle 
in the presence of a magnetic field, and $\varphi$ is the Aharonov-Bohm (A-B) 
phase for a tunneling loop, given by the usual formula ($\phi_0$ is the electron's flux
quantum):
\begin{equation}  
\varphi=2\pi\frac{\Phi}{\Phi_0}, \qquad \Phi_0=\frac{h}{Q}=\phi_0\frac{e}{Q},
\qquad {\rm where} \quad \phi_0=\frac{h}{e}
\label{ABphase}
\end{equation}  
$\Phi_0$ being the appropriate flux quantum ($h$ is Planck's constant) and 
$\Phi={\bf B}\cdot{\bf S}_{\triangle}$ the magnetic flux threading the area
$S_{\triangle}$ offered by the tunneling paths of the ``particle'' in this simple
quantum-mechanical model. The energy asymmetries $E_1, E_2, E_3$ typically 
enter through their combination $D\equiv\sqrt{E_1^2+E_2^2+E_3^2}$. 
One can easily convince oneself that if such a multi-welled potential is used with 
the standard parameter distribution for the STM, 
${\cal P}_{3LS}(\{E_i\},D_0)=\bar{P}/D_0$ 
(the analogous of ${\cal P}_{2LS}(\Delta,\Delta_0)=\bar{P}/\Delta_0$ popular 
for the 2LS), for the description of the generic TS parameters' distribution, one 
would then obtain essentially the same physics as for the STM 2LS-description. In 
other words, there would be no need to complicate the minimal 2LS description in 
order to study glasses at low temperatures, unless structural heterogeneities of the 
RER type and a magnetic field are present. Without the RERs, the interference from 
separate tunneling ``particle'' paths  is likely to give rise but to a very weak 
A-B effect. Hence, it will be those TS nesting in the RERs' voids that will give an 
enhanced A-B effect and these TS can be minimally described though Hamiltonian 
(\ref{3lsmagtunneling}) and with the new parameters' distribution:
\begin{equation}
{\cal P}_{3LS}^*(E_1,E_2,E_3;D_0)=\frac{P^*}{D_0(E_1^2+E_2^2+E_3^2)}
\label{fullnd3lsdistribution}
\end{equation}
enforcing the quasi-ordered character of the RER internal structure 
\cite{KAT2007,Kaw2010}. It is at this 
point, in fact, that the scenario for glasses envisioned by Lebedev \cite{Leb1940}, 
Porai-Koshits \cite{VPK1936}, Greaves \cite{Gre1985}, Gaskell \cite{Gas1998},  
Wright \cite{Wri2014} and Bakai \cite{Bak1994,Bak2013} (as well as the PI's own 
\cite{Jug2018,Jug2004}, of course) gets to be implemented. The new, magnetic-
sensitive TS are named ATS (Anomalous Tunneling Systems) and are not 2LS (the
ATS have a robust collective nature, while the 2LS are almost exclusively atomic). 
We remark that the incipient ``crystallinity'' of the RER \cite{KAT2007,Kaw2010}
calls for near-degeneracy 
in $E_1, E_2, E_3$ simultaneously and not for a single one of them, hence the 
correlated form of (\ref{fullnd3lsdistribution}). Other descriptions, with full 
tetrahedral four-welled potentials for the TS nested in the RER are possible and 
lead to the same physics \cite{Bon2016} as from Eqs. (\ref{3lsmagtunneling}) and 
(\ref{fullnd3lsdistribution}) above. It is important to remark that a distribution 
${\cal P}_{3LS}^*(E_1,E_2,E_3;D_0)$ imposing sharp zero values for the 
asymmetries $\{E_i\}$ (that is, enforcing $E_1^2+E_2^2+E_3^2$=0) would 
lead to no magnetic effects at all. Hence these effects are not consequences of 
true crystallinity, but only of a local ordering-enhancement (such as in the 
cybotactic-grouping vision of melt-quenched glasses or in the RER-scenario).

The next important consideration is that the ATS appear to be rather diluted 
entities in the glass, hence the tunneling ``particles'' are embedded in a rather 
complex charged-particle medium. This embedding however means that the rest of 
the material takes a role in the tunneling of the ATS effective ``particle'', which is 
not moving in a simple vacuum. Sussmann \cite{Sus1962} has shown that this leads 
to local trapping potentials that (for the case of triangular and tetrahedral 
{\it perfect} symmetry) must be characterised by a degenerate ground state. This 
means that as a consequence of the embedding and of the many omitted 
contributions in Eq. (\ref{micrH}) to be replaced by Eq. (\ref{effH}), our minimal 
model (\ref{3lsmagtunneling}) must be chosen with a positive tunneling parameter:
\begin{equation}
D_0>0,
\label{degeneracy}
\end{equation}
where of course degeneracy is always removed by weak disorder in the asymmetries.
Such choice for the tunneling parameter goes completely unnoticed in the 2LS STM.
With this description, the spectrum of the energy levels of a single ATS (supposed
to be associated with the $N_{tunn}$ atomic units active on each distorted 
tetrahedron's face) has the form reproduced in Fig. 7(c) (left panel, PA).  
The intrinsic near-degeneracy of (\ref{fullnd3lsdistribution}) implies that this
model can be used in its $D/D_0\ll 1$ limit, so long as the magnetic field is ``weak''
(i.e. $\varphi\ll 1$) which in turn produces for low enough temperatures an
{\em effective} magnetic-field dependent 2LS and greatly simplifies the analysis
together with the limit $\varphi\to 0$ which can be used for relatively ``weak'' 
magnetic fields. The above Extended Tunneling Model (ETM) consists then in a
collection of independent, non-interacting 2LS described by the STM and 2LS-like
non-interacting 3LS of internal collective nature described by 
Eqs. (\ref{3lsmagtunneling}) and (\ref{fullnd3lsdistribution}) above and in the 
said $D/D_0\ll 1$ and $\varphi\to 0$ limits. The 3LSs are quasi-particles nested 
in the voids between the RER and the magnetic-field insensitive 2LS are distributed 
in the RER and (as it turns out) overwhelmingly around their contact interfaces (see 
Section F). In the said limits, a reasonable approximate form for the lower energy gap is 
of the simple form \cite{Jug2004} $E=\sqrt{D^2+D_0^2\varphi^2}$; for higher fields 
this can be easily corrected (see below) still to make a manageble analytic theory. 

At this point, the contribution to the magnetisation from the tunneling currents located
(initially, at least) on the faces of the distorted tetrahedra (one tetrahedron per each 
RER ``void'') can be calculated from the ETM above described, using standard quantum 
statistical mechanics. As already shown in \cite{Bon2015}, in the limit of weak fields 
the following expression can be used (for a given ${\bf B}$-direction):
\begin{eqnarray}
M_{intr}^{tunn}(T,B)&=& \pi P^{*} n_{ATS}\frac{1}{B} \Big\{
\int_{E_{c1}}^{E_{c1}} dE 
\tanh\Big( \frac{E}{2k_BT} \Big) \ln\Big( 
\frac{E^2-D_{0min}^2\varphi^2}{D_{min}^2} \Big)  \cr
&+&  \int_{E_{c2}}^{\infty} dE \tanh\Big( \frac{E}{2k_BT} \Big) 
\ln\Big( \frac{E^2-D_{0min}^2\varphi^2}{E^2-D_{0max}^2\varphi^2} \Big) \Big\}
\label{ATSmagnet}
\end{eqnarray}
where: $D_{min}$, $D_{0min}$ and $D_{0max}$ are material-dependent cutoffs
of the ATS parameter distribution, Eq. (\ref{fullnd3lsdistribution}); 
$E_{c1}=\sqrt{D_{min}^2+D_{0min}\varphi^2}$ and 
$E_{c2}=\sqrt{D_{min}^2+D_{0max}\varphi^2}$ are special points in the density 
of states of the ATS spectrum; finally $n_{ATS}$ is the ATS concentration (quantity 
always convoluted with $P^*$, the single ATS energy-distribution parameter which 
when normalized gives the somewhat undetermined relationship 
$1/P^*=2\pi\ln(\frac{D_{0max}}{D_{0max}})\ln(\frac{E_{max}}{D_{min}})$, 
$E_{max}$ being linked to the (unknown) single ATS-well attempt frequency: 
in practice, $P^*$ is of order $1/2\pi$). 
Eq. (\ref{ATSmagnet}) needs to be averaged over ATS orientations with respect to
${\bf B}$. When the magnetic field is not so weak (this is determined by a material-
dependent characteristic field $B^*$ \cite{Pal2011,Bon2015}) the above expression
can be cured by simply shifting:
\begin{equation}
\varphi^2\to\varphi^2\big( 1-\frac{1}{27}\varphi^2 \big)
\label{shift}
\end{equation}
which corrects the $1/B$ prefactor into $\big( 1-\frac{1}{54}(B/B^*)^2 \big)/B$
when averaging over ATS triangular path's geometrical orientation is carried out. 

More details in a future publication. However, we anticipate here the unusual 
temperature behaviour of this formula in the weak field limit, which completes the 
Curie formula, for the susceptibility $\chi_{intr}^{tunn}$ due to the contribution to the 
bulk magnetisation $M$ coming from the ATS in the ``voids'' between the jammed 
RER. Since the effective-TLS approximation for the magnetic ATS is applicable, we 
get in the end:
\begin{equation}
 \chi_{intr}^{tunn}(T)=\frac{2}{3} \pi^3 \frac{n_{ATS}P^*}{\phi_0^2}
\Big( \frac{Q}{e} \Big)^2 S_{\triangle}^2 (D_{0max}^2-D_{0min}^2)
\int_{D_{min}}^{\infty} \frac{dE}{E^2} \tanh\Big( \frac{E}{2k_BT} \Big)
\label{notCurie}
\end{equation}
and this is not quite Curie's law. In fact, working out the limits we have:
\begin{eqnarray}
\lim_{T\gg D_{min}/k_BT} \chi_{intr}^{tunn}(T)&=&
\chi_{intr}^{tunn}(T)=\frac{2}{6} \pi^3 \frac{n_{ATS}P^*}{\phi_0^2}
\Big( \frac{Q}{e} \Big)^2 S_{\triangle}^2 (D_{0max}^2-D_{0min}^2)\frac{1}{k_BT}
\cr
\lim_{T\ll D_{min}/k_BT} \chi_{intr}^{tunn}(T)&=&
\chi_{intr}^{tunn}(T)=\frac{2}{3} \pi^3 \frac{n_{ATS}P^*}{\phi_0^2}
\Big( \frac{Q}{e} \Big)^2 S_{\triangle}^2 (D_{0max}^2-D_{0min}^2)\frac{1}{D_{min}}
\label{nonC-limits}
\end{eqnarray}
so that at $T=0$ this becomes a constant, while at slightly higher temperatures a 
fake Curie-law type term is recovered. This is part of the novel behaviour at high values 
of $1/T$. In the next Sections we shall see that, indeed, the data for $M$ at weak fields 
do show deviations from the $1/T$ Curie law precisely as indicated by 
Eq. (\ref{nonC-limits}). The fitting formula for the suceptibility of a piece of glass then
becomes of the form:
\begin{equation}
\chi=-\chi_L+\frac{C}{T}+G\int_{D_{min}}^\infty \frac{dE}{E^2} \tanh\Big( \frac{E}
{2k_BT} \Big)
\label{susceptibility}
\end{equation}
where $\chi_L$ is the closed-shells' Larmor contribution, $C\propto n$(Fe) is the Curie
constant and $G\propto n_{ATS}$ is the new constant due to the collective coherent 
tunneling currents in the glass structure. More about the use of these formulas in 
Section D.

\begin{figure}[!h]
\centering
{ \vskip -1cm
\includegraphics[scale=0.4]  {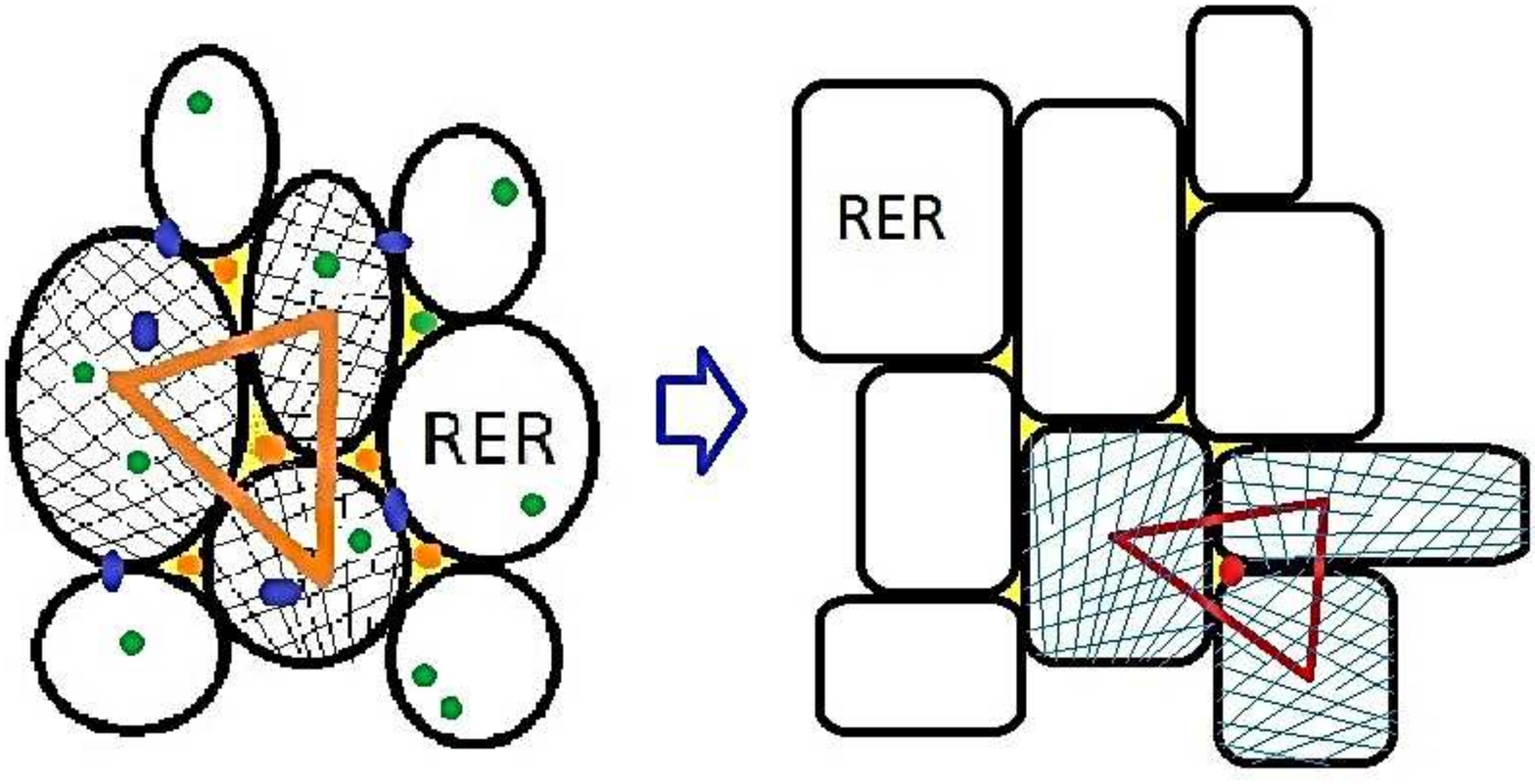}  
}
\vskip -0cm
\caption{ Compactification upon cooling of the RER close-packed arrangement 
(going from $T\simeq T_g$, left, to $T\to$0, right hand side). The RER grow at
the lowest temperatures at the expense of the material in the void, increasing at the 
same time the area of the RER-RER interfaces. Defects, in going from l.h.s to r.h.s., 
have been removed.
}
\label{compact}
\end{figure}

\vskip 5mm
The last important ingredient for this theory is the substantial $T$-dependence that
was found for $N_{tunn}$, the number of atomic-scale TS that make up, collectively,
a single ATS quasi-particle described above. This crucial quantity clearly enters into 
the ATS renormalized tunneling parameters \cite{Jug2013,theory}: the effective 
charge $Q=N_{tunn}e$, the flux-threaded surface 
$S_{\triangle}\propto N_{tunn} a_0^2$ ($a_0$ is of the order of Bohr's radius, see 
also below) and also the tunneling amplitude $D_0\propto N_{tunn}\Delta_{0min}$ 
and energy gap's cutoff $D_{min}\propto N_{tunn} \Delta_{min}$ ($\Delta_{0min}$ 
and $\Delta_{min}$ being the atomic 2LS's STM cutoffs) \cite{Jug2013,Pal2011}. This 
state of affairs is responsible for the large ATS cutoffs-combinations values always 
found in data-fitting the ETM to available experimental data (a situation fully confirmed 
in this work, see Tables \ref{BK7parametersH} and \ref{BK7parametersT}). 
The temperature-dependence of $N_{tunn}$ arises from
the shrinking of the RER tetrahedral void space upon deep cooling, when the RER 
walls grow at the expense of the surviving melt material in the voids, extending the
RER-RER interfacial area and creating a tight cellular structure as shown in 
Fig. \ref{compact}. This dependence can be written down, on general 
thermodynamics grounds, in the form
\begin{equation}
N_{tunn}(T)=N_0\exp\{-\frac{\Delta\mu}{k_BT}\}
\label{Ntunnequil}
\end{equation}
where $\Delta\mu=\mu_{melt}-\mu_{RER}$ is the chemical potential difference
between the active melt in the void and the RER surface. We will use this expression
because it introduces one fit parameter less. However, a more precise non-equilibrium 
evaluation can be made by considering the RER surface growth, by activated 
deposition of melt species from the void's space (which shrinks when temperature 
decreases). Then one gets:
\begin{equation}
N_{tunn}(T)=N_0\Big( 1-\Gamma\exp\{\frac{E_0}{k_BT}\} \Big)
\label{Ntunnnonequil}
\end{equation}
where $E_0$ is an activation energy and $\Gamma$ a geometric parameter. As it
turns out, both expressions (\ref{Ntunnequil}) and (\ref{Ntunnnonequil}) give very 
similar fitting values for $\Delta\mu$ and $E_0$ and same behaviour for 
$N_{tunn}/N_0$ down to the range of $T$ where new, unexpected phenomena 
appear.  The expression Eq. (\ref{Ntunnequil}) then allows to relate the ETM best-fit 
parameters to a reference temperature ($T_0$=1.26 K, due to alternative knowledge 
of the parameters from $C_p$ fits in that range \cite{Bon2015}). 

\vskip 1.0cm

%----------------------------------------------------------------------------------------
%	SECTION 3(Pt. 1)
%----------------------------------------------------------------------------------------

{\bf C. Some More Data Fitting Results.} 

\begin{figure}[!h]
\centering
{ \vskip -0cm
   \subfigure[]{\includegraphics[scale=0.5]  {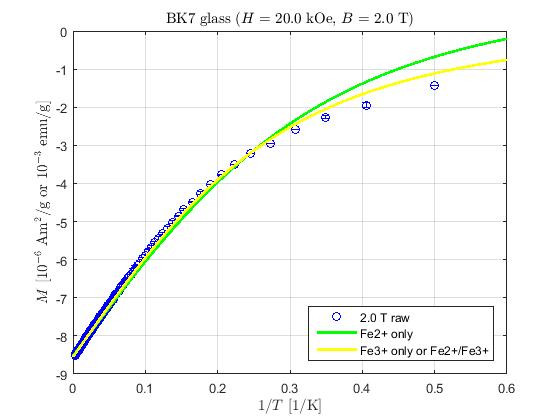} } 
% [scale=0.20] {BK7_specimen.jpg}
 \vskip -0cm
   \subfigure[]{\includegraphics[scale=0.5] {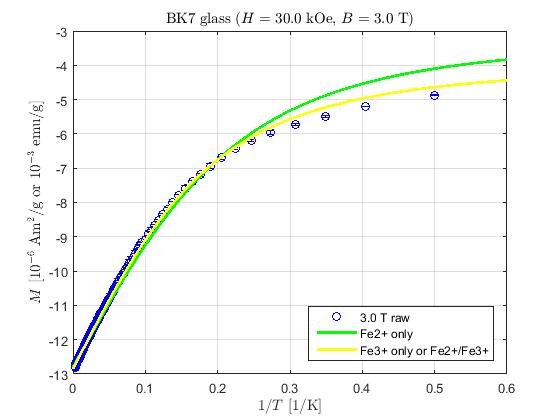} }
}
\caption{ 
(a) BK7 glass sample's measurement of the magnetisation $M$ at $H$=20.0 kOe (2.0 T)
in a SQUID-magnetometer. Raw data (black dots, with errorbars) being best fitted with
Eq. (1) in the PA, so with one paramagnetic species, or two (in which case the 
Fe$^{3+}$-only scenario is always selected. 
(b) Same as in (a), but for $H$=30.0 kOe (3.0 T).
}
\label{BK7LLfits}
\end{figure}

{\bf C1) BK7 glass.} 
Fig. \ref{BK7LLfits} reports two typical fit attempts with the LL-formula (Eq. (1) in the
PA) for our SQUID-magnetisation data from a BK7-glass chip sample, much in the spirit 
of Fig. 1 in the PA (which was for Duran). It is apparent that the LL form does not fit 
the data that well also for BK7 and especially at low $T$, moreover it induces the 
wrong conclusion that the sample contains only Fe$^{3+}$ (which is false). The 
contribution unaccounted for comes from the intrinsic magnetisation 
$M_{intr}^{tunn}(T,H)$ discussed in the previous Section B. 

\begin{figure}[!h]
\centering
{ \vskip -0cm
\includegraphics[scale=0.6]  {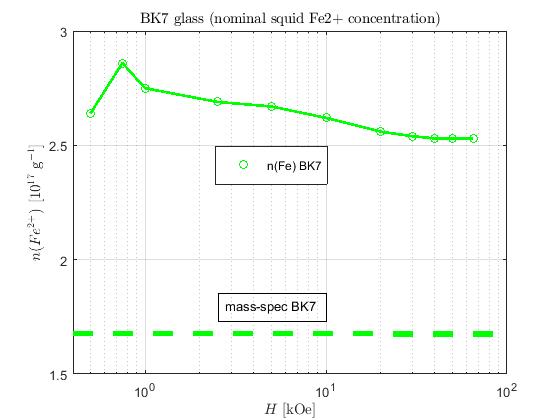}  
}
\vskip -0cm
\caption{ The dependence of the Fe$^{2+}$ concentration extracted from 
SQUID-runs $M$ vs. $T$ and as a function of $H$. The shape of the ''intrinsic'' 
$\Delta n=n$(Fe2+)-$n_{MS}$ curve is very reminiscent of other magnetic effects 
in glasses \cite{Jug2018} with a sharp maximum at low $H$ followed by a slow
decrease. The dashed line is the $n_{MS}$(Fe) mass-spec value for Fe in BK7.
}
\label{BK7Fe3conc}
\end{figure}

Tables \ref{BK7LLH} and \ref{BK7LLT} (which we reproduce here from the PA for 
convenience's sake) as well as Fig. \ref{BK7Fe3conc} show for BK7-glass how 
important the fluctuations are in both $n$(Fe) and $\chi_L$ by SQUID-scanning in 
$T$ or $H$ while keeping $H$ or $T$ constant, respectively, and employing the 
LL-fit form, Eq. (1) in the PA, alone. In Fig. \ref{BK7Fe3conc} only the dependence 
of the putative $n$(Fe2+) is shown since mostly Fe$^{2+}$ is expected (and found, 
in the proposed theory) and because the $H$-dependence of the missing contribution 
($\Delta n=n_{put}$(Fe2+)-$n_{MS}(Fe)$) coming from the glass is very reminiscent 
of other magnetic-field effects in glasses \cite{Jug2018}. 

\begin{table}[!hbp]
\begin{center}
%\hspace*{-2cm}
\begin{tabular}{ |c|c|c|c|c|c|c|c|c|c|c| } % 11-column LL analysis, fixed H

\hline

Magnetic Field (kOe) & 1.0 & 2.5 & 5.0 & 10.0 & 20.0 & 30.0 & 40.0 & 50.0 & 
 65.0 & mass-spec \\
\hline
BK7 LL-parameters &  &  &  &  &  &  &  &  &  &  \\
\hline

$n$(Fe$^{3+}$) 10$^{17}$ g$^{-1}$ & 1.888 & 1.849 & 1.845 & 1.849 & 1.867 &
1.888 & 1.906 & 1.914 & 1.931 & 1.657 $\pm$ 0.019 \\ 

$\chi_L$ 10$^{-7}$ emu/gOe & 3.922 & 4.064 & 4.139 & 4.257 & 4.277 & 4.283 &
4.295  & 4.294 & 4.297 & - \\

\hline

$n$(Fe$^{2+}$) 10$^{17}$ g$^{-1}$ & 2.752 & 2.690 & 2.667 & 2.621 & 2.559 &
2.541 & 2.536 & 2.526 & 2.526 & 1.657 $\pm$ 0.019 \\

$\chi_L$ 10$^{-7}$ emu/gOe & 3.922 & 4.064 & 4.136 & 4.249 & 4.262 & 4.266 &
4.277 & 4.276 & 4.278 & - \\

\hline

\end{tabular}
\caption{ LL-fitting (Eq. (1) in PA) parameters extracted from different SQUID runs 
of $M$ vs. $1/T$ at stated $H$-values for a BK7-prism chip having mass 96.8 mg. 
The first raw is obtained from best fits when both Fe$^{3+}$ and Fe$^{2+}$ 
species are allowed (in practice one always obtains $n$(Fe$^{2+}$)=0); for 
comparison, the results when Fe$^{2+}$-only is allowed are shown in the second 
raw. The precise mass-spec value of $n$(Fe) is also indicated (errorbar was estimated
from spectrometer specifications, repeating the analysis and by dissolving separate 
BK7-prism chips, including this one). }
 \label{BK7LLH}
\end{center}
\end{table}

\begin{table}[!htp]
\begin{center}
%\hspace*{-2cm}
\begin{tabular}{ |c|c|c|c|c|c|c|} % 7-column LL analysis, fixed T

\hline

Temperature (K) & 2.0 & 4.5 & 7.5 & 10.0 & 20.0 & mass-spec \\
\hline
BK7 LL-parameters &  &  &  &  &  &  \\
\hline

$n$(Fe$^{3+}$) 10$^{17}$ g$^{-1}$ & 1.746 & 1.877 & 1.969 & 2.035 & 3.135 &
1.657 $\pm$ 0.019 \\ 

$\chi_L$ 10$^{-7}$ emu/gOe & 4.178 & 4.269 & 4.326 & 4.350 & 4.686 & - \\

\hline

$n$(Fe$^{2+}$) 10$^{17}$ g$^{-1}$ & 2.414 & 2.905 & 3.358 & 3.681 & 6.206 &
1.657 $\pm$ 0.019 \\

$\chi_L$ 10$^{-7}$ emu/gOe & 4.178 & 4.269 & 4.326 & 4.350 & 4.686 & - \\

\hline

\end{tabular}
\caption{ The same as in Table \ref{BK7LLH}, but as extracted from SQUID-runs of 
$M$ vs. $H$ at stated $T$-values. }
 \label{BK7LLT}
\end{center}
\end{table}

\begin{figure}[!h]
\centering
{ \vskip -0cm
\includegraphics[scale=0.6]  {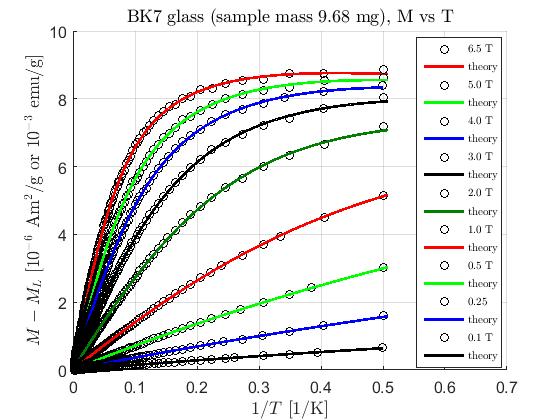}  
}
\vskip -0cm
\caption{ The original SQUID data $M$ vs. $1/T$ (after shifting the raw data by 
a constant $M_L=-\chi_L H$ term with $\chi_L$ as given in the Table 
\ref{BK7parametersH} below) and their best fit to the ETM-theory. The best-fitting 
procedure has been always applied to the raw data.
}
\label{BK7HFig}
\end{figure}

\begin{figure}[!h]
\centering
{ \vskip -0cm
\includegraphics[scale=0.6]  {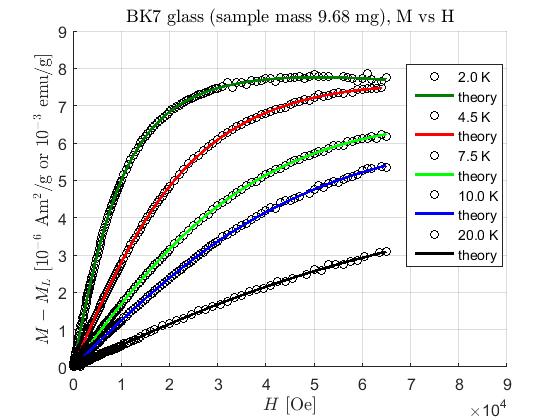}  
}
\vskip -0cm
\caption{ The original SQUID data $M$ vs. $H$ (after shifting the raw data by 
a $M_L=-\chi_L H$ term with $\chi_L$ as given in the Table \ref{BK7parametersT} 
below) and their best fit to the ETM-theory. The best fit was applied to the raw data.
}
\label{BK7TFig}
\end{figure}

In Fig.s \ref{BK7HFig} and \ref{BK7TFig} the SQUID-runs that were collected for 
BK7 are reported, for fixed $H$ and $T$ respectively, with the best fits of the data 
using $M(T,H)=M_{LL}(H/T)+M_{intr}^{tunn}(T,H)$ as described in the previous 
Section of these SI ($M_{LL}(H/T)$ as given by Eq. (1) in the PA). 

The material and ETM parameters extracted from the best fit to the raw data are 
reported in Table \ref{BK7parametersH} for SQUID-runs in fixed magnetic field. 

\begin{table}[!htbp]
\begin{center}
%\hspace*{-2cm}
\begin{tabular}{ |c|c|c|c|c|c| } % 6-column (2 moot) full-fit analysis, fixed H

\hline
\hline

Constant-$H$ & & BK7 glass   & mass-spec & & BK7 glass  \\

parameters    &  & this sample & this sample & &  2000 sample 
\cite{Her2000,Bon2015} \\

\hline
\hline

$n$(Fe$^{2+}$) 10$^{17}$ g$^{-1}$ &   & 1.099 $\pm$ 0.001  &   &   & 0.067  \\ 
$n$(Fe$^{3+}$) 10$^{17}$ g$^{-1}$ &   & 0.599 $\pm$ 0.001  &   &   & 0.034  \\

\hline
$n_{tot}$(Fe) 10$^{17}$ g$^{-1}$     &    & 1.698 $\pm$ 0.001  & 1.657 $\pm$ 0.019 &   & 0.101  \\

$\chi_L$ 10$^{-7}$ emu/gOe              &   & 4.227 $\pm$ 0.095 &     &   & 3.600  \\

\hline
\hline

$n_{ATS}$ 10$^{16}$ g$^{-1}$         &    & 1.352 $\pm$ 0.001 &    &    & 1.400  \\
$D_{min}$ 10$^{-2}$ K                      &    & 2.671 $\pm$ 0.076 &    &    & 5.990  \\
$D_{0min}\frac{Q}{e}S_{\triangle}$  10$^5$ K\AA$^2$ &    & 2.717 $\pm$ 0.310 &    &    & 0.887  \\
$D_{0max}\frac{Q}{e}S_{\triangle}$ 10$^5$ K\AA$^2$  &    & 2.925 $\pm$ 0.297 &    &    & 1.200  \\
$\Delta\mu$ K                                      &   & 0.319 $\pm$ 0.067 &    &     & 0.480 \\

\hline
\hline

\end{tabular}
\caption{ Table of parameters extracted from best-fitting with 
$M=M_{LL}+M_{intr}^{tunn}$  (Eq. (1) in the PA and Eq. (\ref{ATSmagnet})) our 
$M$ vs. $T$ SQUID-runs gathered in Fig.  (\ref{BK7HFig}). The set on the 
left-hand-side is for our own BK7 sample chip (96.8 mg mass), while the set of 
parameters on the right is for the best-fit \cite{Bon2015} to SQUID-magnetisation 
data published in \cite{Her2000} for yet another BK7 sample in $H$=30.0 kOe (mass 
and mass-spec analysis unknown). 
}
\label{BK7parametersH}
\end{center}
\end{table}

They are to be compared with the parameters in Table \ref{BK7parametersT} for 
SQUID-runs at fixed temperature.

\begin{table}[!htbp]
\begin{center}
%\hspace*{-2cm}
\begin{tabular}{ |c|c|c|c|c|c| } % 6-column (2 moot) full-fit analysis, fixed H

\hline
\hline

Constant-$T$ & & BK7 glass   & mass-spec & & BK7 glass  \\

parameters    &  & this sample & this sample & &  2000 sample 
\cite{Her2000,Bon2015} \\

\hline
\hline

$n$(Fe$^{2+}$) 10$^{17}$ g$^{-1}$ &   & 1.098 $\pm$ 0.001  &   &   & 0.067  \\ 
$n$(Fe$^{3+}$) 10$^{17}$ g$^{-1}$ &   & 0.569 $\pm$ 0.001  &   &   & 0.034  \\

\hline
$n_{tot}$(Fe) 10$^{17}$ g$^{-1}$     &    & 1.667 $\pm$ 0.001  & 1.657 $\pm$ 0.019 &   & 0.101  \\

$\chi_L$ 10$^{-7}$ emu/gOe              &   & 4.137 $\pm$ 0.004 &     &   & 3.600  \\

\hline
\hline

$n_{ATS}$ 10$^{16}$ g$^{-1}$         &    & 1.340 $\pm$ 0.001 &    &    & 1.400  \\
$D_{min}$ 10$^{-2}$ K                      &    & 5.238 $\pm$ 0.488 &    &    & 5.990  \\
$D_{0min}\frac{Q}{e}S_{\triangle}$  10$^5$ K\AA$^2$ &    & 2.107 $\pm$ 0.032 &    &    & 0.887  \\
$D_{0max}\frac{Q}{e}S_{\triangle}$ 10$^5$ K\AA$^2$  &    & 2.160 $\pm$ 0.033 &    &    & 1.200  \\
$\Delta\mu$ K                                      &   & 0.615 $\pm$ 0.615 &    &     & 0.480 \\

\hline
\hline

\end{tabular}
\caption{ The same as in Table \ref{BK7parametersH}, but for our $M$ vs. $H$ 
SQUID-runs gathered in Fig.  (\ref{BK7TFig}). The set on the left-hand-side is for our 
own BK7 sample chip (96.8 mg mass), while the set of parameters on the right is, 
for the sake of comparison, for the best-fit \cite{Bon2015} to SQUID-magnetisation 
data ($M$ vs. $T$ at $H$=30.0 kOe) published in \cite{Her2000} for yet another BK7 
sample (mass and mass-spec analysis unknown). 
}
\label{BK7parametersT}
\end{center}
\end{table}
 
Looking at these extracted parameters, Tables \ref{BK7parametersH} and
\ref{BK7parametersT},  the following immediate conclusions can be drawn: 
(a) The extracted total-Fe concentrations are in excellent agreement with the
value extracted from mass-spectrometry; the small discrepancy (Table 
\ref{BK7parametersH}) can be attributed to the presence of minority paramagnetic
Ti- and Cr-species in BK7 that have been lumped into Fe-only Langevin terms (the 
purpose is here to show that theory can also predict Fe-concentrations from SQUID
runs as accurate as mass-spec's while keeping the number of fit-parameters low). 
The tiny discrepancy (if any) between the fixed-$H$ value of $\chi_L$ and the 
fixed-$T$ value remains however unexplained. 
(b) Although the 2000 BK7 sample was much cleaner in Fe-contents, 
the ratio Fe$^{2+}$/Fe$^{3+}$ ($\approx$ 1.93 for our sample, 1.97 for the 
2000 sample) remains the same: BK7 has $\approx$ 66\% Fe$^{2+}$ and 33\%
Fe$^{3+}$ and this seems to be linked to oxide-composition. (c) Likewise for the
concentration of ATS, that appears to be the same in both samples; we expect,
however, that $n_{ATS}$ is also strongly linked to cooling-rate and other 
vetrification conditions and the size 2$\xi$ of BK7's RER remains the same as for the 
2000 sample, $\approx$ 62 nm. (d) The other best-fit parameters for the ETM 
remain only of the same order of magnitude for the two BK7 samples: the ETM is, 
after all, just an effective theory and can be improved.

\vskip 1.0cm

%----------------------------------------------------------------------------------------
%	SECTION 3(Pt. 2)
%----------------------------------------------------------------------------------------

\begin{figure}[!h]
\centering
{ \vskip -0cm
   \subfigure[]{\includegraphics[scale=0.5]  {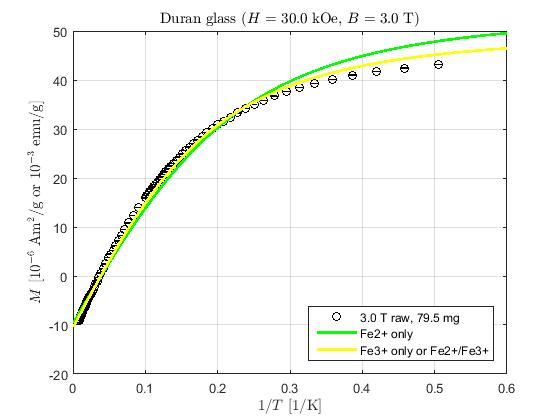} } 

 \vskip -0cm
   \subfigure[]{\includegraphics[scale=0.5] {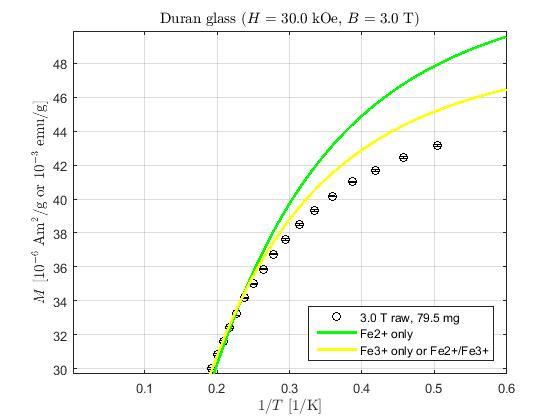} }
}
\caption{ 
(a)  Duran 79.5 mg, 30.0 kOe (3.0 T) raw data fitted with the LL form Eq. (1) in the PA.
(b) Same as (a), but for the lowest temperatures in an expanded vertical scale, to show
the strong deviation between naive LL theory and the data.
}
\label{Duran3TLLfits}
\end{figure}

\begin{figure}[!h]
\centering
{ \vskip -0cm
   \subfigure[]{\includegraphics[scale=0.5]  {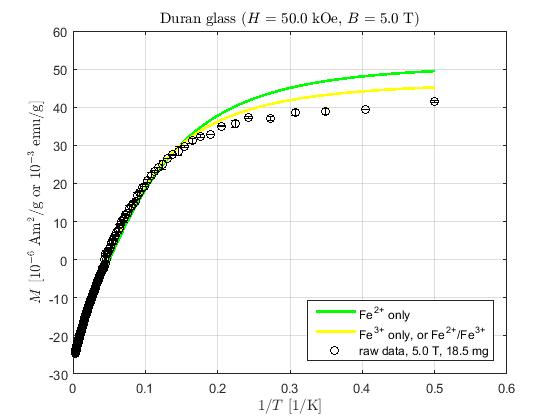} } 
% [scale=0.20] {BK7_specimen.jpg}
 \vskip -0cm
   \subfigure[]{\includegraphics[scale=0.5] {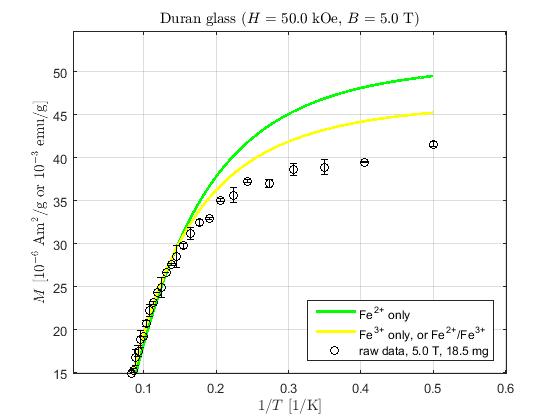} }
}
\caption{ 
(a)  Duran 18.5 mg, 50.0 kOe (5.0 T) raw data fitted with the LL form Eq. (1) in the PA
and as already reported in Fig. 1 of the PA.
(b) Same as (a), but for the lowest temperatures in an expanded vertical scale, to show
the strong deviation between naive LL theory and the data.
}
\label{Duran5TLLfits}
\end{figure}

{\bf C2) Duran and BAS glasses.} 
In Figs. \ref{Duran3TLLfits} and \ref{Duran5TLLfits} 
we report the raw data and naive LL SQUID-magnetisation theory fits
for the Duran shards we examined and for $H$=30.0 kOe (3.0 T) and 50.0 kOe (5.0 T)
(caution: different samples). It is clear (especially from the details at low $T$) that a
new (albeit small) contribution is absent in the fitting theoretical expression. 

\begin{figure}[!h]
\centering
{ \vskip -0cm
   \subfigure[]{\includegraphics[scale=0.5]  {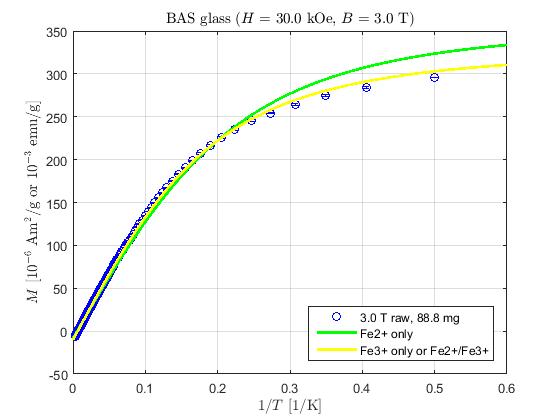} } 
 \vskip -0cm
   \subfigure[]{\includegraphics[scale=0.5] {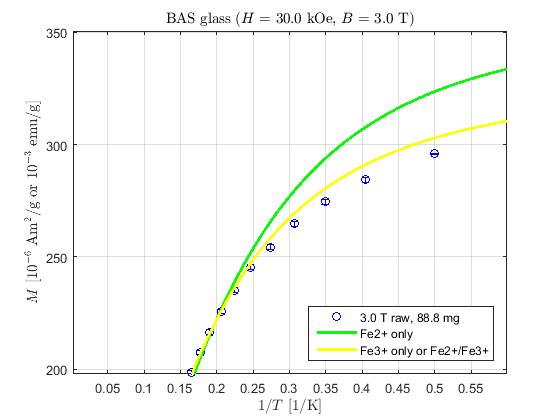} }
}
\caption{ 
(a)  BASw 88.8 mg, 30.0 kOe (3.0 T) raw data fitted with the LL form Eq. (1) in the PA.
(b) Same as (a), but for the lowest temperatures in an expanded vertical scale, to show
the strong deviation between naive LL theory and the data.
}
\label{BAS3TLLfits}
\end{figure}

\begin{figure}[!h]
\centering
{ \vskip -0cm
   \subfigure[]{\includegraphics[scale=0.5]  {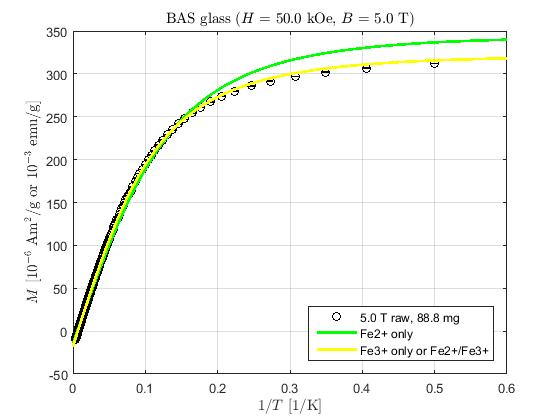} } 
 \vskip -0cm
   \subfigure[]{\includegraphics[scale=0.5] {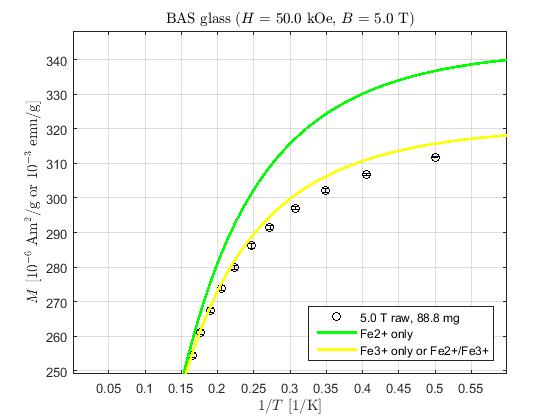} }
}
\caption{ 
(a)  BASw 88.8 mg, 50.0 kOe (5.0 T) raw data fitted with the LL form Eq. (1) in the PA
and as already reported in Fig. 1 of the PA.
(b) Same as (a), but for the lowest temperatures in an expanded vertical scale, to show
the strong deviation between naive LL theory and the data.
}
\label{BAS5TLLfits}
\end{figure}

In Figs. \ref{BAS3TLLfits} and \ref{BAS5TLLfits} we report the raw data and naive LL 
SQUID-magnetisation theory fits for the BAS-glass fragments we examined and for 
$H$=30.0 kOe (3.0 T) and 50.0 kOe (5.0 T). Again, although the Fe-content is much
higher, it is clear (especially from the details at low $T$) that a new (albeit small) 
contribution is absent in the fitting theoretical expression. 

\begin{figure}[!h]
\centering
{ \vskip -0cm
\includegraphics[scale=0.3]  {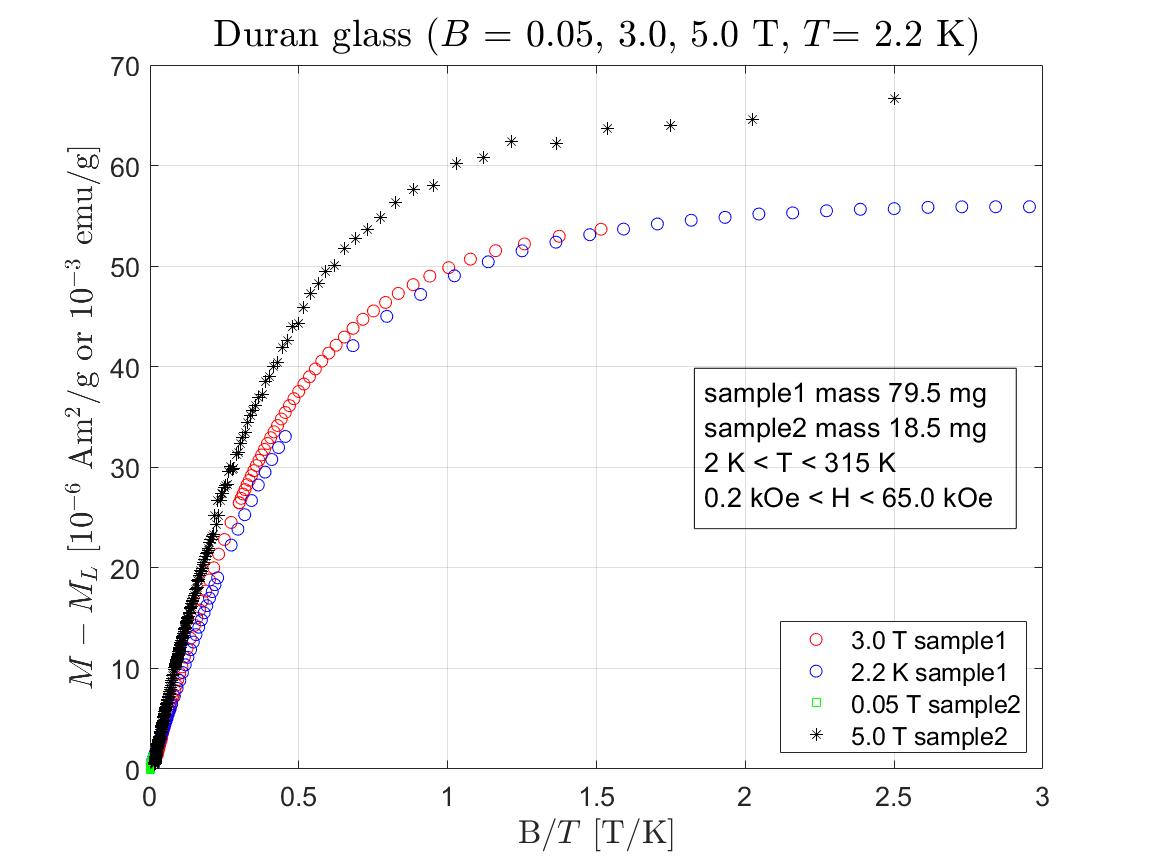}  
}
\vskip -0cm
\caption{ Universality test for our SQUID-run measurements on Duran glass (all what
was available). The raw data have been re-plotted in the same spirit as for Fig. 2 in 
the PA. As always, $B=\mu_0 H$.
}
\label{Duran-universal}
\end{figure}

\begin{figure}[!h]
\centering
{ \vskip -0cm
\includegraphics[scale=0.6]  {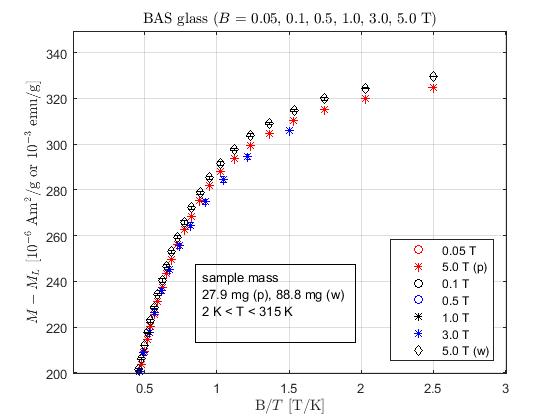}  
}
\vskip -0cm
\caption{ Universality test for our SQUID-run measurements on BAS glass (for clarity, not 
all of our fixed-$H$ runs are reported). The raw data have been re-plotted in the same 
spirit as for Fig. 2 in the PA. $B=\mu_0 H$.
}
\label{BAS-universal}
\end{figure}

In Fig.s  \ref{Duran-universal} and \ref{BAS-universal} we report the universality tests 
for $M-M_L$ as a function of $H/T$ for these other two multi-silicate glasses that we 
have studied. Again, we subtracted the Larmor contribution $-\chi_L H$ as determined 
from the naive $M_{LL}$ fit from Eq. (1) in the PA. Clearly, universality does not hold. 

We now report, for the Duran and BAS glasses, the qualitative MS analysis for the 
Fe-group elements. Again, Fe is the predominant element and the others are present in
such smaller concentration and have the wrong $J_s$ value to affect our conclusions.

\begin{table}[!htp]
\begin{center}
%\hspace*{-2cm}
\begin{tabular}{ |c|c|c|c|c|c|c|c| } % 8-column qual-MS analysis, BK7 98.6 mg

\hline

{\bf Fe} & Ti & V & Cr & Mn & Co & Ni & Cu \\

\hline

{\bf 1} & 0 & 0.005 & 0.017 & 0.023 & 0.002 & 0.004 & 0.008 \\

\hline

\end{tabular}
\caption{ Relative concentration (taking $n$(Fe)=1) of all other Fe-group magnetic
elements present in similar Duran-glass shards as those SQUID-characterised in this work 
and from our own qualitative MS analysis. }
 \label{qual-MS-Duran}
\end{center}
\end{table}

\begin{table}[!htp]
\begin{center}
%\hspace*{-2cm}
\begin{tabular}{ |c|c|c|c|c|c|c|c| } % 8-column qual-MS analysis, BK7 98.6 mg

\hline

{\bf Fe} & Ti & V & Cr & Mn & Co & Ni & Cu \\

\hline

{\bf 1} & 0 & 0.001 & 0.011 & 0.005 & 0 & 0.005 & 0.023 \\

\hline

\end{tabular}
\caption{ Same as previous Table \ref{qual-MS-Duran}, but for BAS-glass calcinated paste 
scrapings.  }
 \label{qual-MS-BAS}
\end{center}
\end{table}

For comparison, we report in Table \ref{qual-MS-BK7} the results already presented 
in the PA of the qualitative MS analysis for the BK7 glass samples that were studied
in the SQUID magnetometer.

\begin{table}[!htp]
\begin{center}
%\hspace*{-2cm}
\begin{tabular}{ |c|c|c|c|c|c|c|c| } % 8-column qual-MS analysis, BK7 98.6 mg

\hline

{\bf Fe} & Ti & V & Cr & Mn & Co & Ni & Cu \\

\hline

{\bf 1} & 0.1 & 0 & 0.1 & 0.01 & 0 & 0.01 & 0.01 \\

\hline

\end{tabular}
\caption{ Same as previous Tables \ref{qual-MS-Duran} and \ref{qual-MS-BAS}, but 
for the very same BK7-chips SQUID-characterised in this work. }
 \label{qual-MS-BK7}
\end{center}
\end{table}

\begin{figure}[!h]
\centering
{ \vskip -0cm
\includegraphics[scale=0.3]  {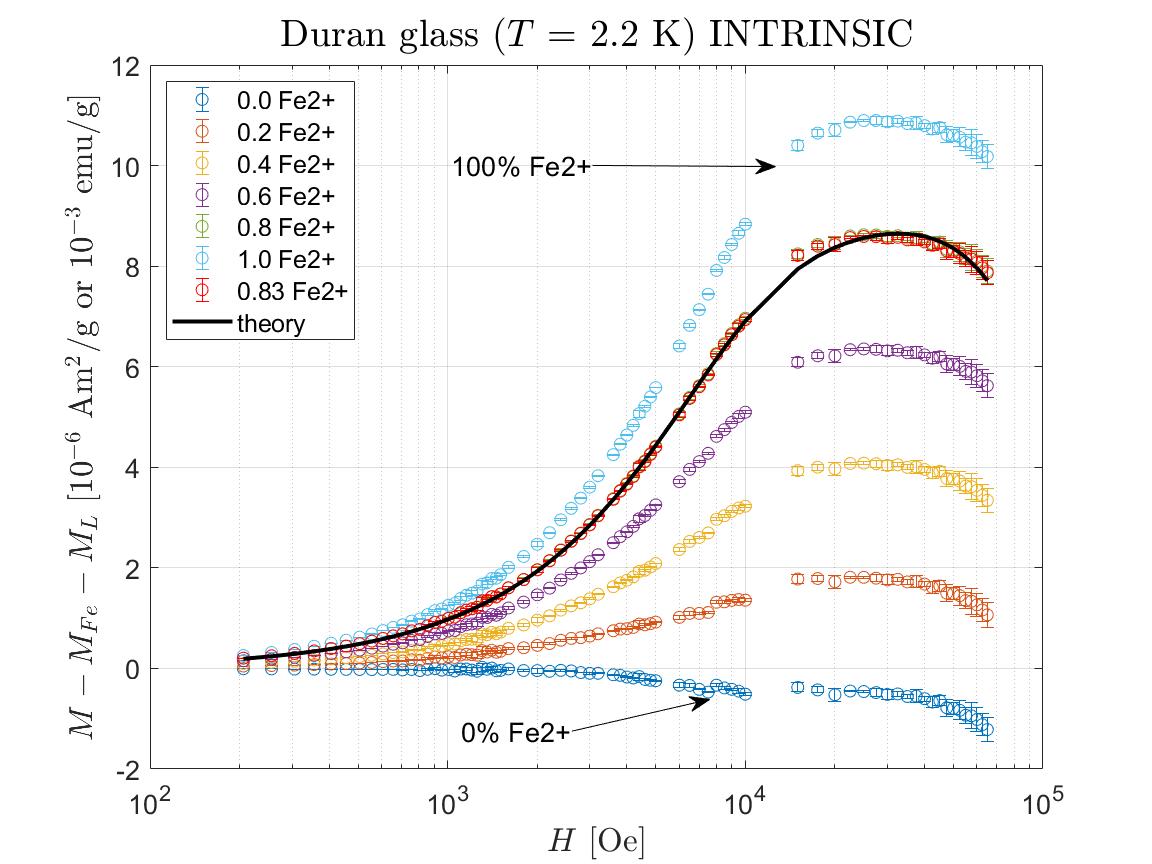}  
}
\vskip -0cm
\caption{ Re-plot of raw data for Duran glass after subtraction of the Langevin contribution 
from $x n_{MS}$(Fe) Fe$^{2+}$-ions and $(1-x) n_{MS}$(Fe) Fe$^{3+}$-ions with 
$n_{MS}$(Fe)=1.47$\times$10$^{18}$ g$^{-1}$ the MS-ascertained value. The raw data 
have been re-plotted in the same spirit as for Fig. 4(a) in the PA. From top to bottom
$x$=1.0, 0.8, 0.6, 0.4, 0.2, 0.0 and the ETM theory singles out the right concentration
ratio.
}
\label{DuranHx}
\end{figure}

\begin{figure}[!h]
\centering
{ \vskip -0cm
\includegraphics[scale=0.3]  {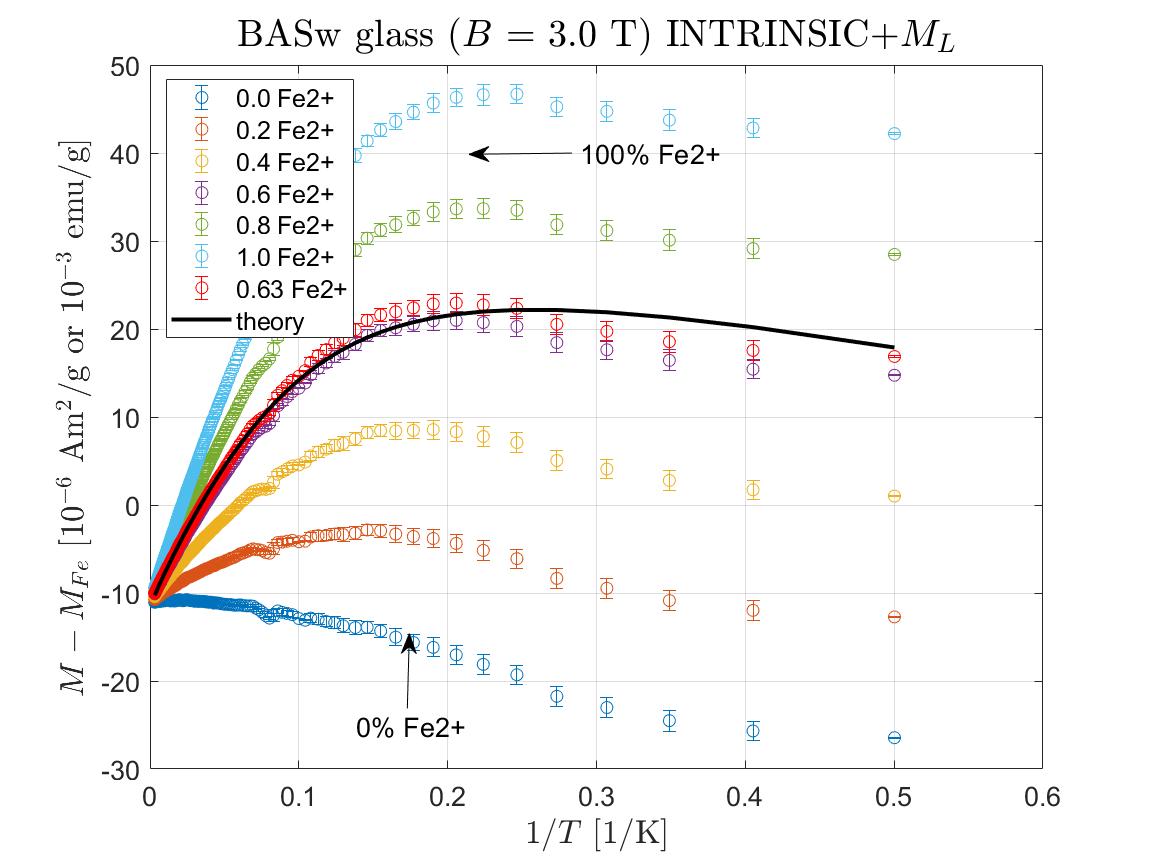}  
}
\vskip -0cm
\caption{ As for Fig. \ref{DuranHx}, but for BAS-w glass and with
$n_{MS}$(Fe)=7.13$\times$10$^{18}$ g$^{-1}$ the MS-ascertained value of 
Fe-concentration. The raw data have been re-plotted in the same spirit as for 
Fig. 3(a) in the PA. 
}
\label{BASTx}
\end{figure}

More data and data analyses for the SQUID measurements on the Duran- and BAS-glasses
will be presented elsewhere \cite{JCP2021}. Here we conclude by presenting two further 
studies of the effect of subtracting fixed fractions $x$ of Fe$^{2+}$ and $1-x$ of
Fe$^{3+}$ Langevin contributions from the raw data for $M-M_L$ (the latter $M_L$
as determined from naive $M_{LL}$ best-fits). What remains is not only non-zero, but 
indicates that it presents the odd shape as a function of either $1/T$ or $H$ as predicted
by the ETM theory and practically for any value of $x$. In Fig. \ref{DuranHx} for Duran- 
at a temperature of 2.2 K kOe and in Fig. \ref{BASTx} for BAS-glass in a field of 30.0 kOe 
(3.0 T) we see that the application of the ETM theory singles out the right (MS-ascertained)
values of $x$ for both Duran- and BAS-glass and give good fits to the modified experimental
data.

\begin{figure}[!h]
\centering
{ \vskip -0cm
   \subfigure[]{\includegraphics[scale=0.25]  {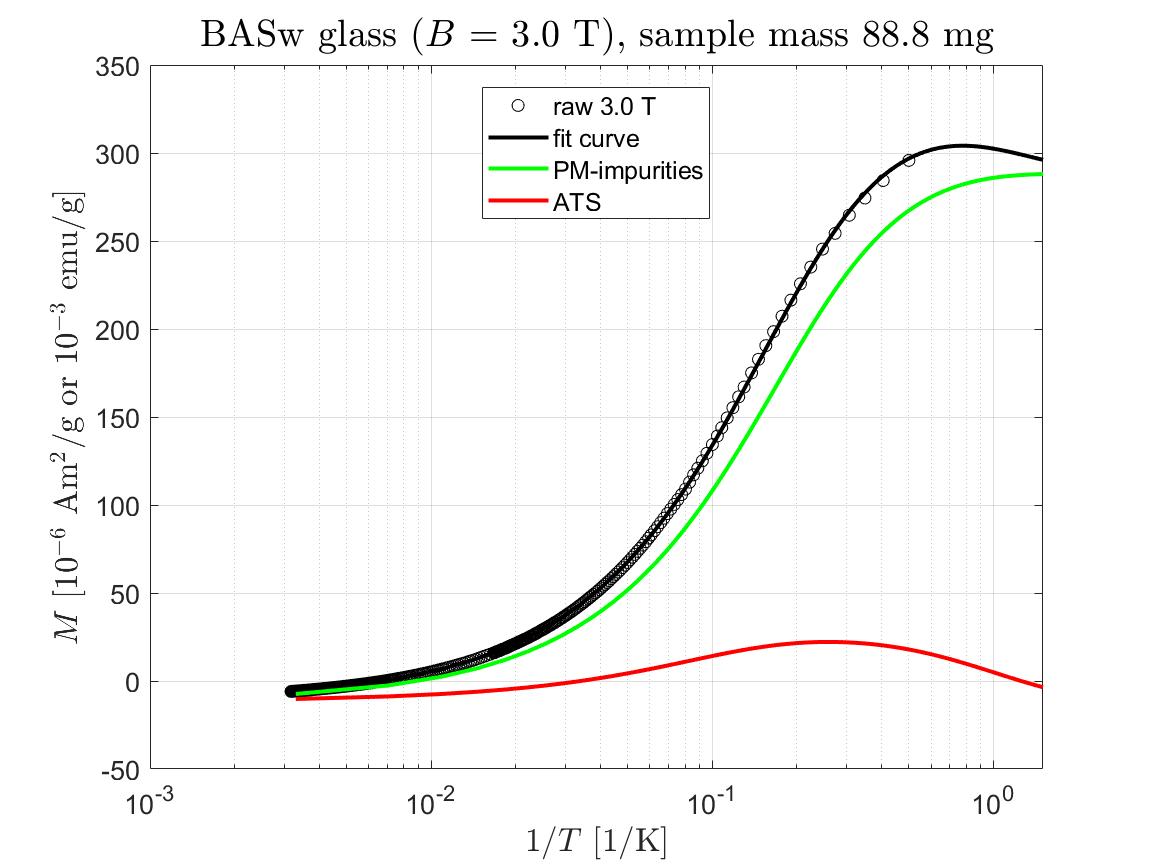} } 
 \vskip -0cm
   \subfigure[]{\includegraphics[scale=0.25] {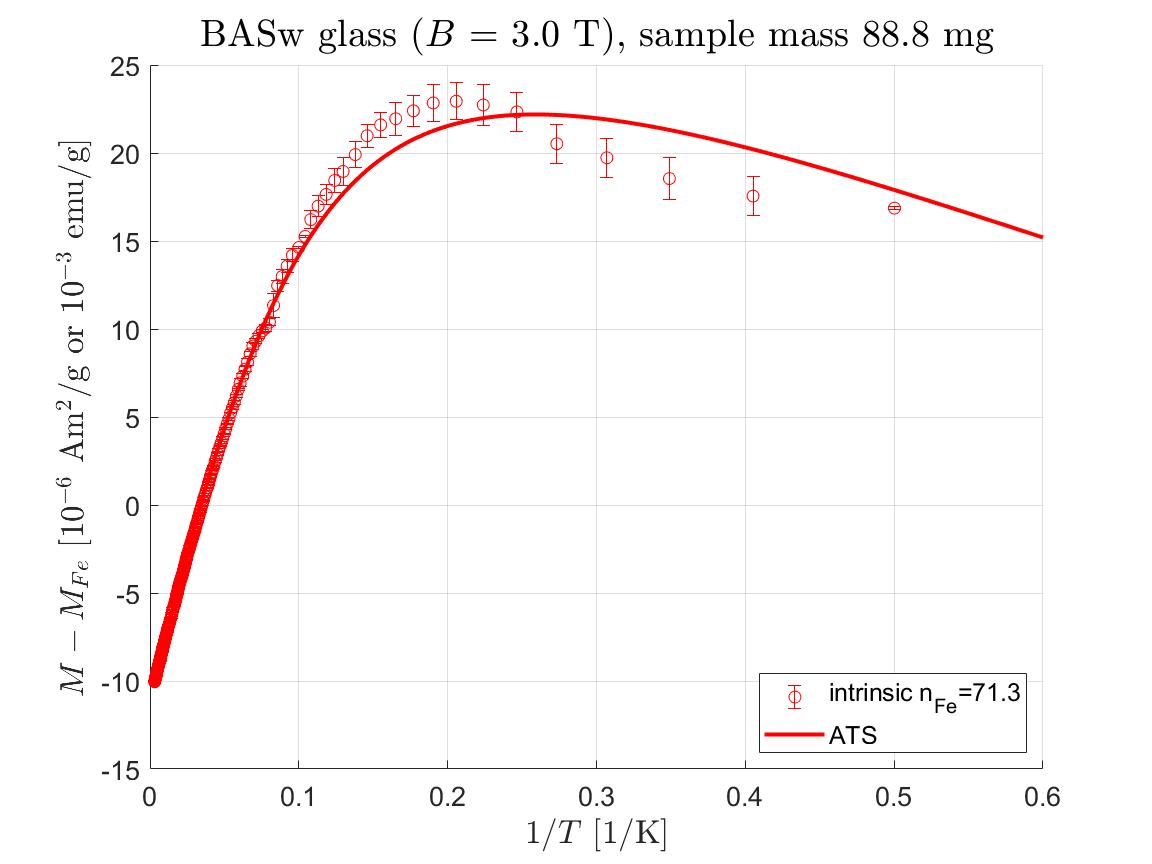} }
}
\caption{ 
(a)  BASw 88.8 mg, 30.0 kOe (3.0 T) raw data (black circles) fitted with the LL 
form Eq. (1) in the PA (green curve) where the Fe concentration is fixed at the 
MS-acertained value (71.3 $\times$ 10$^{17}$ g$^{-1}$). The Fe(2+)/Fe(3+)
partition is fixed by the present theory ($x$=0.63) but any other value of $x$
gives worsening LL agreement with the data. The missing contribution is here
indicated in red (from theory) and provides the full fit in black to the raw data.
(b) Same as (a) for the missing contribution to the raw data, now as subtracted
data points (red dots) with the LL contributions taken away. The red continuous
curve is the same as in (a), the ETM theory's ATS contribution. 
}
\label{BAS3T}
\end{figure}

In order to clearly state the case for glass paramagnetism, in Fig. \ref{BAS3T}(a) 
we show the raw magnetisation data for a case at fixed $H$ (30.0 kOe or 3.0 T) 
along with the Larmor-Langevin contributions consistent with the MS-ascertained 
Fe-concentration and optimal Fe(2+)/Fe(3+) partitioning in a sample of BAS-w 
glass. The $n$(Fe$^{2+}$) is from the theory best-fit (the value of $\chi_L$ also), 
however any other partitioning would give a curve that cannot explain the data. 
The missing contribution is accounted for only by the ETM and is the contribution 
from the ATS, also shown in the figure (Fig. \ref{BAS3T}(a)). The theoretical 
ATS curve clearly shows a broad peak as a function of $1/T$ at the lowest 
temperatures. For the same sample and measurements, in Fig. \ref{BAS3T}(b) 
we report the subtracted data points for the intrinsic $M_{intr}$ part as a function
of $1/T$ and the theoretical ATS contribution for comparison. The theory fit is 
satisfactory, but new phenomena (yet unaccounted for by theory) appear at the 
lowest reached temperatures.

%----------------------------------------------------------------------------------------
%	SECTION 4
%----------------------------------------------------------------------------------------

\vskip 1.0cm

\begin{figure}[!h]
\centering
{ \vskip -0cm
   \subfigure[]{\includegraphics[scale=0.3]  {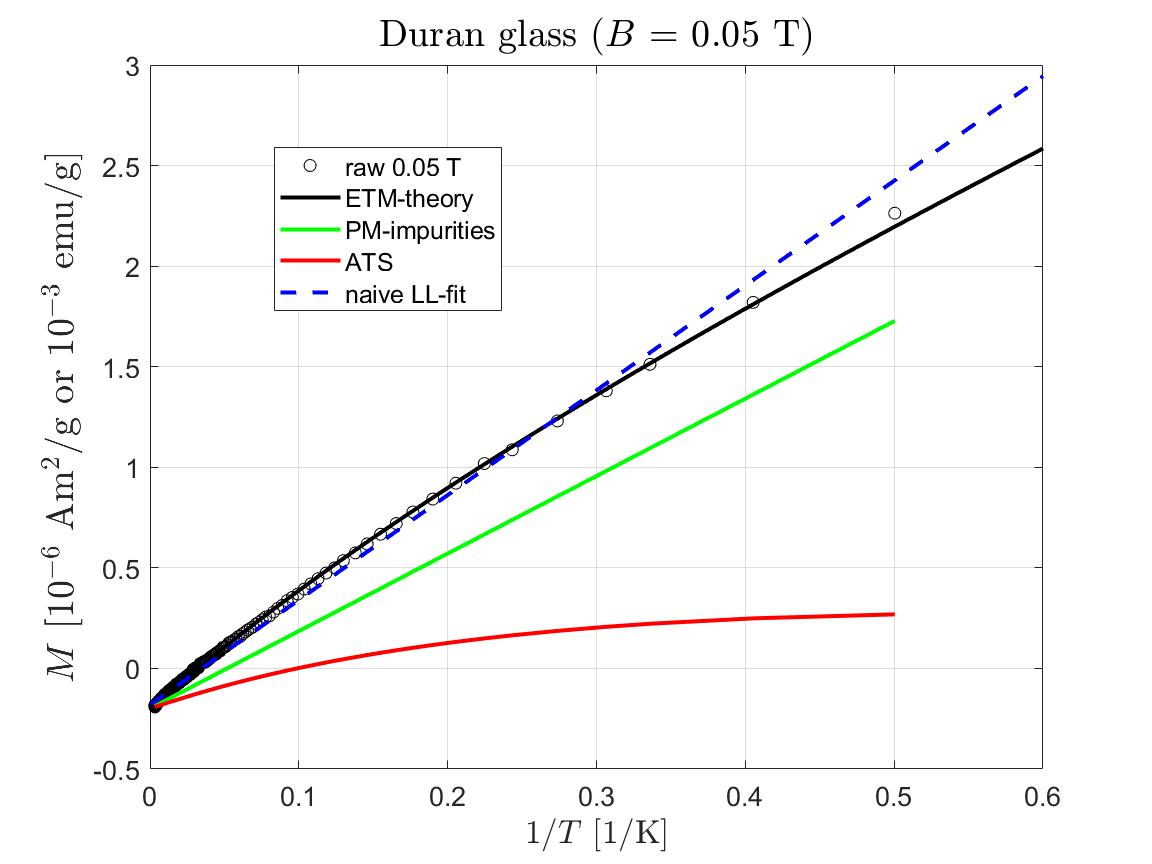} } 
 \vskip -0cm
   \subfigure[]{\includegraphics[scale=0.3] {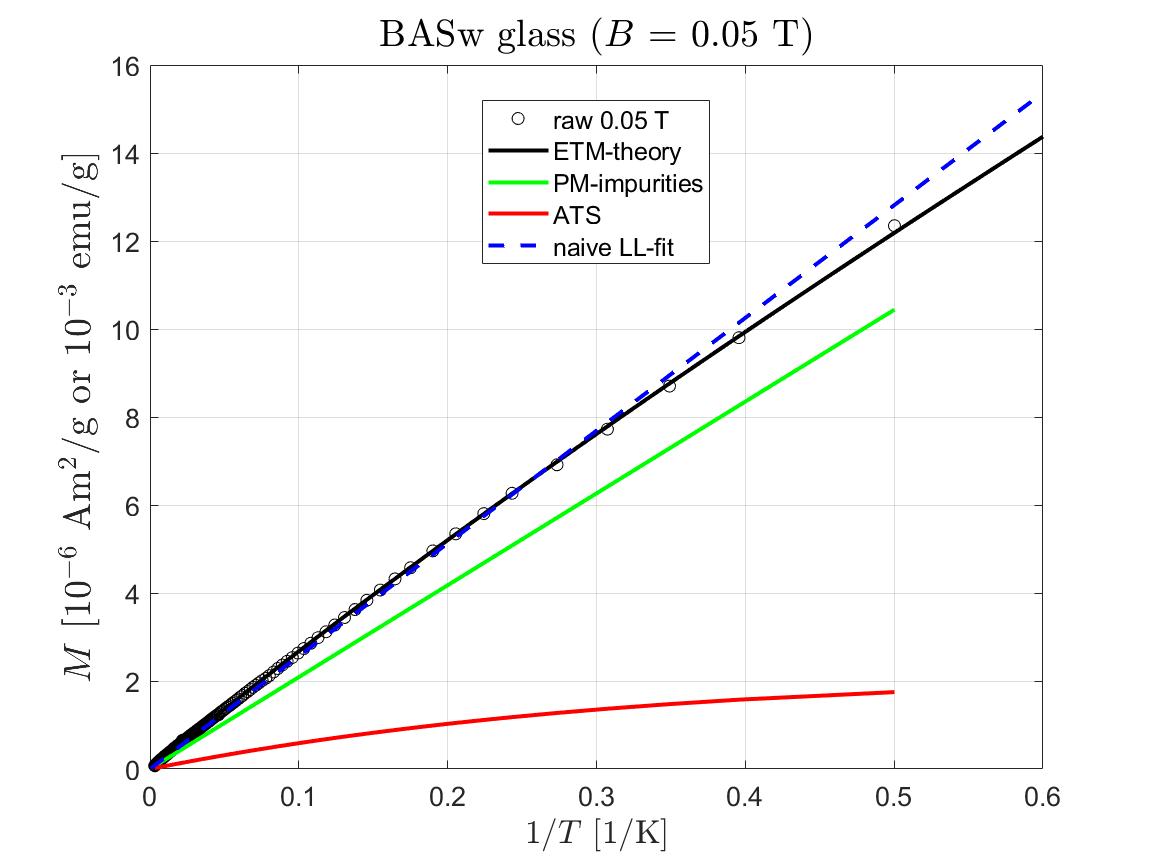} }
}
\caption{ 
The failure of the Curie law in glasses. (a) Data for the magnetisation of a small shard
of Duran-glass: The dashed blue line is the naive Curie law, which extracts the wrong 
value of the Fe-concentration (and all Fe$^{3+}$, which is false), the real Langevin 
contribution (Fe$^{2+}$ and Fe$^{3+}$) being the green line. While the red line is the 
contribution from the ATS coherent-tunneling currents, the black line is the full best fit.
(b) Same as (a), but for a BAS-glass sample. Errorbars are about the same size of the dots. 
}
\label{lowH}
\end{figure}

{\bf D. Further Discussion about Curie Law and  Comparison with ETM-Theory. }
The experimental data for the bulk magnetisation of point-like samples of the 
above-stated glassy systems have been analysed with the formula:
\begin{equation}
M(T,H)=-\chi_L H + M_{{\rm Fe}}(H/T) + M_{intr}^{tunn}(H,T) 
\label{fullmag}
\end{equation}
where the first (Larmor) and the second (Langevin) terms are those of Eq. (1) in the PA
and the third is the ATS coherent-tunneling currents' ensuing magnetisation as discussed 
in Section B (briefly), Eq. (\ref{ATSmagnet}), and in Ref. \cite{Bon2015}. In the limit of 
weak magnetic fields $H/T\to 0$ the second term and the third term assume the forms
given in Eq. (\ref{susceptibility}) for the susceptibility $M/H$. 
Here, we want to show how the form Eq. (\ref{susceptibility}) (alternatively the full form 
containing Eq. (\ref{ATSmagnet})) can account for deviations from the Curie law.
Fig. \ref{lowH}(a) for Duran- and (b) for BAS-glass show our weakest magnetic field
($H$=500.0 Oe, or 0.05 T) bulk magnetisation data points, with the dashed blue line 
representing the Curie law. Not only this procedure extracts a faulty $n$(Fe) concentration 
value, but also it does not provide a good fit at the lowest temperatures. The fit
to the data by our theory (full black line) considerably improves agreement at all values of 
$1/T$, moreover it extracts the right concentration $n$(Fe) and works out (green line)
the correct Langevin contribution. The red line represents the contribution from the ATS
coherent-tunneling currents. 

These considerations suggest an elegant new test for the ETM-, magnetic ATS-theory
presented in Section B. If the Larmor susceptibility $\chi_L$ of the sample can be 
ascertained in an independent way, then measuring the magnetic susceptibility $\chi$
as a function of very low temperatures $T$ (below 4.0 K and at least down to 300 mK)
will produce an interesting graph plotting $T(\chi-\chi_L)$ vs. $T$. In an ordinary 
weakly Fe-doped insulating non-magnetic crystal this should produce a constant $C$, 
while from the expression in Eq. (\ref{susceptibility}) we should have (note: $\chi_L$ 
taken with its negative sign here):
\begin{equation}
T\Big( \chi-\chi_L \Big)=C+GT\int_{D_{min}}^\infty \frac{dE}{E^2} \tanh\Big( \frac{E}
{2k_BT} \Big)  
\label{prediction}
\end{equation}
and this should increase linearly like $C+\frac{G}{D_{min}}T$ with temperature, with 
$C\propto n$(Fe) the Curie constant (one for Fe$^{2+}$ and one for Fe$^{3+}$ in 
fact). A simple form that should be contrasted with a simple constant $C$ for 
$\lim_{T\to 0}T(\chi-\chi_L)$ in the case of the crystal.

%----------------------------------------------------------------------------------------
%	SECTION 5
%----------------------------------------------------------------------------------------

\vskip 1.0cm

\begin{figure}[!h]
\centering
{ \vskip -0cm
   \subfigure[]{\includegraphics[scale=0.32]  {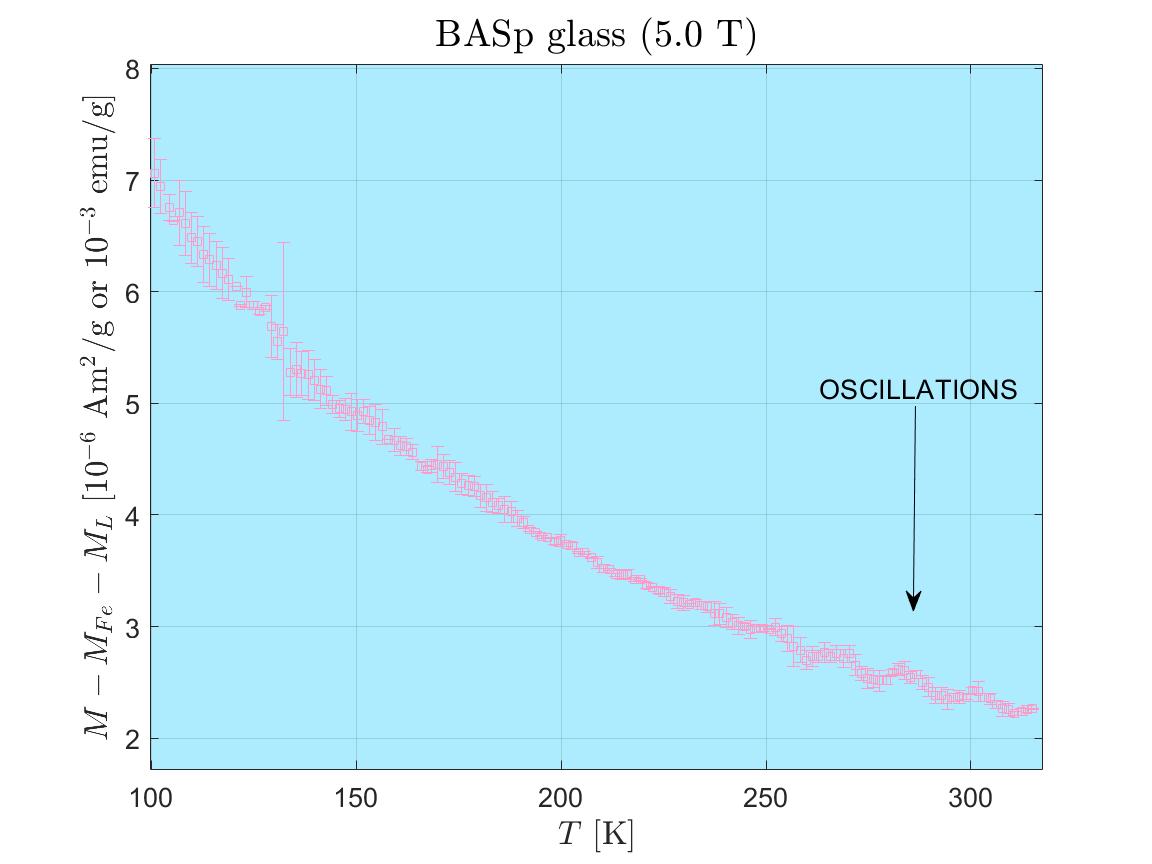} } 
% [scale=0.20] {BK7_specimen.jpg}
 \vskip -0cm
   \subfigure[]{\includegraphics[scale=0.32] {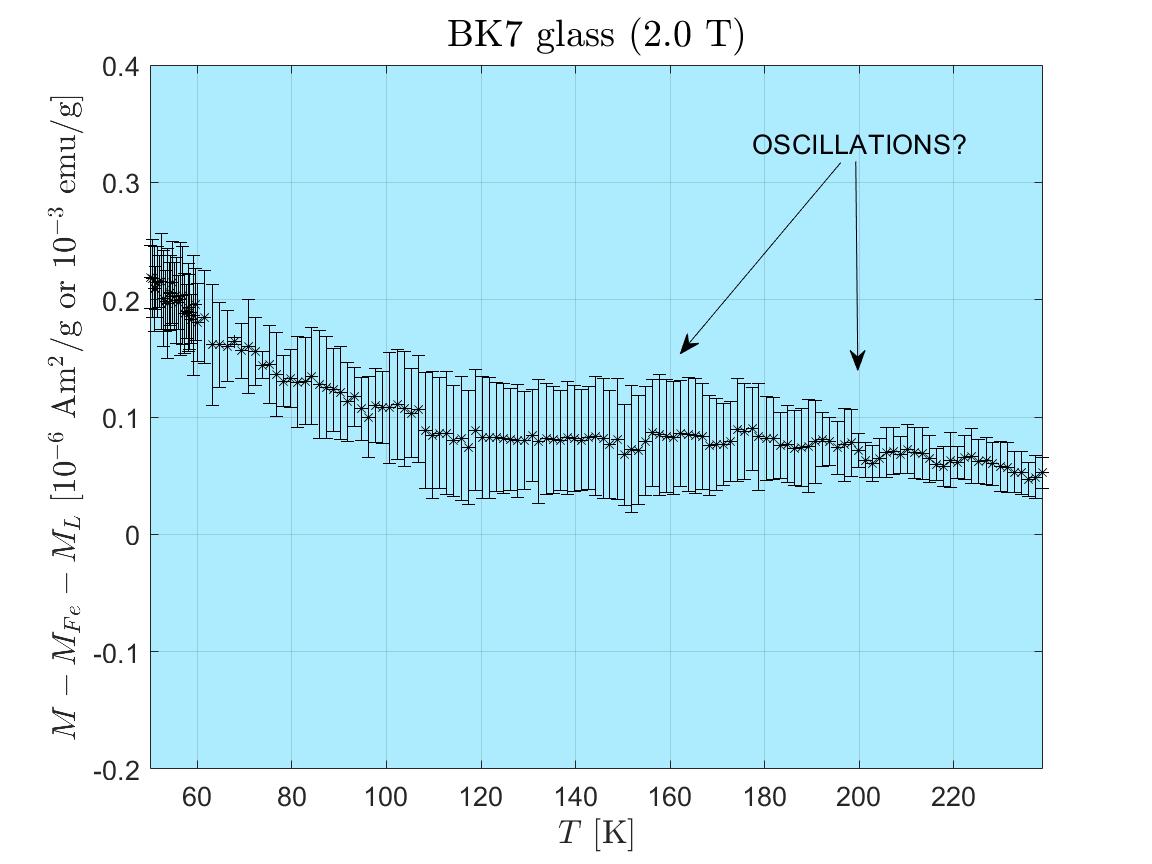} }
}
\caption{ 
(a)  BAS-p glass at 50.0 kOe (5.0 T) raw data after subtraction of the LL form (Eq. (1) in 
the PA) for MS/theory-fit extracted Fe-concentrations: distinct oscillations as a function of 
$T$ are noted, almost as clear as for Duran glass (PA Fig. 6). 
(b) Same as (a) but for BK7 glass at 20.0 kOe (2.0 T): oscillations, if present, are
much weaker for this type of glass, signalling that the RER size is largest for this glass.
}
\label{osc_BASpBK7}
\end{figure}

{\bf E. The $M_{intr}$ Temperature Oscillations and the High Field Limit.} 
Here we comment some more on the theoretical qualitative interpretation of the data, for 
all of the glasses. 
For the oscillations in the graph of $M_{intr}$ vs. $T$ (Fig. 6 in the PA) we offer a 
tentative but very compelling explanation. For a start, the oscillations can be observed
also for the BAS-p magnetisation data, polished of the Larmor-Langevin contributions.
In Fig. \ref{osc_BASpBK7} we reproduce the data for $M_{intr}$ at indicated conditions.   

\begin{figure}[!h]
  \centering
  \vskip -0mm
  \includegraphics[scale=0.08] {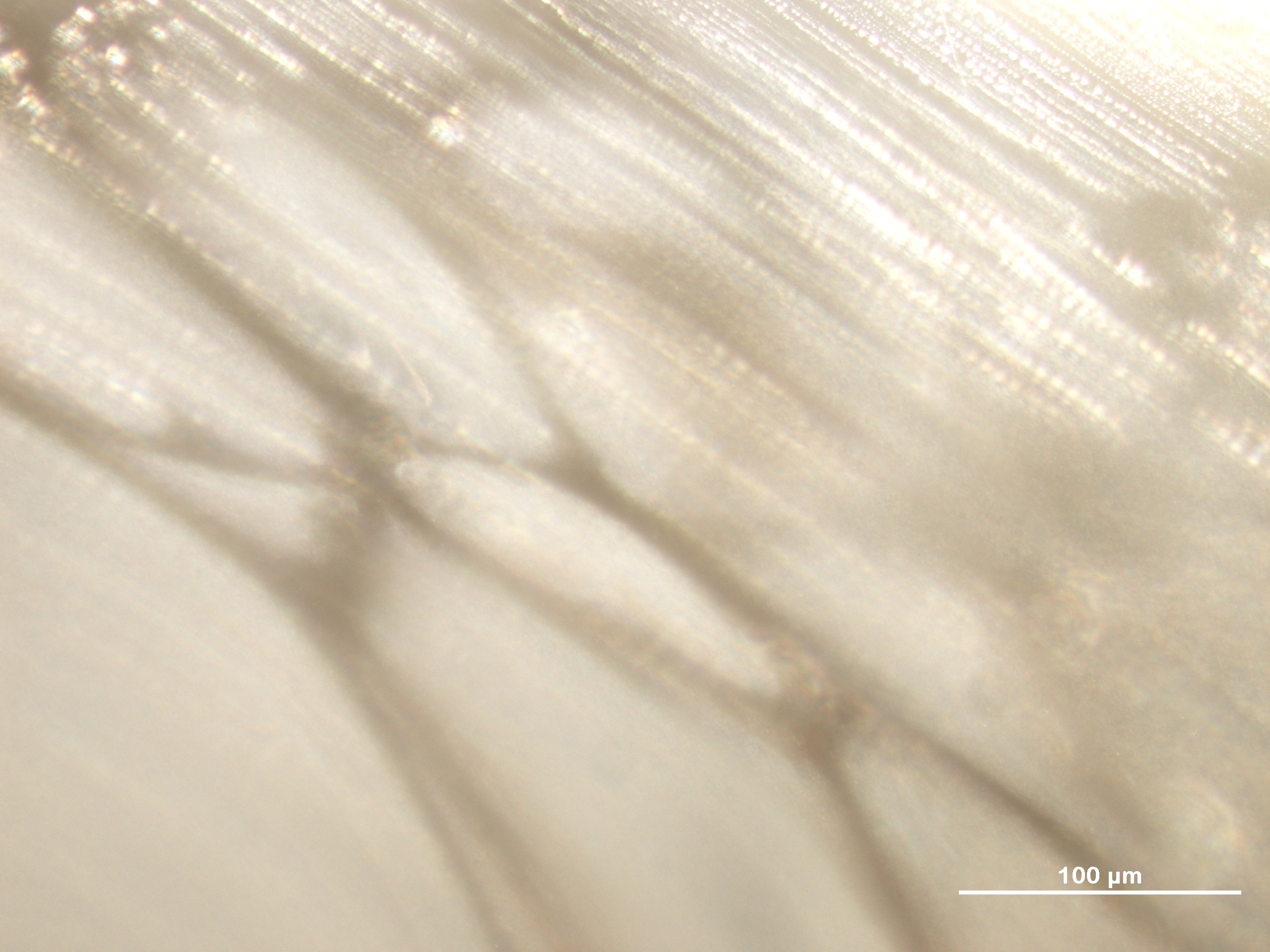}
 \vskip -0mm
\caption{ Ordinary optical microscopy image of the cell-structure of a manufact of 
seeded BASb (black) glass \cite{JSRV2021}, showing the jammed cells medium-range 
structure. Marked self-forming grooves on the clear manufact's surface are also visible. 
}
 \label{BAS1_cells}
\end{figure}

In order to provide a plausible theoretical explanation for these oscillations in $M_{intr}$
we resort to the polycluster or cellular model of glass structure advocated for in this 
paper. We do have in fact some direct experimental evidence for the existence of such 
cells jammed against each other, obtained precisely for BAS glass with an expedient 
(seeding the hot glass-forming liquid with foreign particles of the appropriate size 
\cite{JSRV2021}) which produces order 100 $\mu$m size cells visible with an ordinary 
microscope. Fig. \ref{BAS1_cells} in particular shows the white cells under a microscope 
for one "black" BAS glass manufact of some 10 mm width and 5 mm thickness.
However for unseeded glass -- or self-seeded, or highest-$T_m$ (melting point) component 
seeded glass -- the size of these cells is estimated to be from O(1) to O(100) nm and then
collective magnetic effects are expected. 

\begin{figure}[!h]
\centering
{ \vskip -10mm
\subfigure[]{\includegraphics[scale=0.70]  {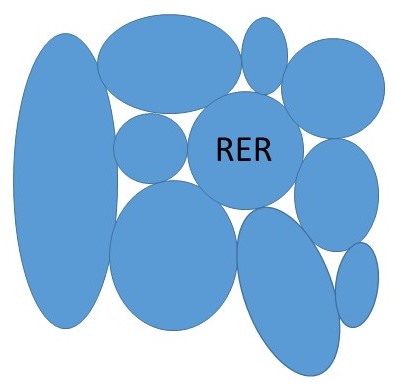}} 
\vskip -0mm
\subfigure[]{\includegraphics[scale=0.70] {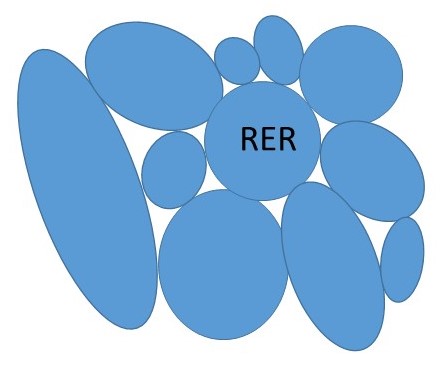} }
}
\caption{ (a) 2D cartoon of a (portion of) medium-range RER-cell structure of our model
glass, at some temperature $T_1$ below $T_g$. The (blue) ovals are better-ordered RER
(solid-like particle regions) and contain (white) in-between liquid-like regions. (b) The 
same as in (a), but for a different, close temperature $T_2$: the RER have moved to a
new equilibrium configuration, changing the RER-RER voids' total perimeter lenght per 
unit area (proportional to the intrinsic magnetisation) slightly.
}
\label{cellsT1T2}
\end{figure}

To simulate this structure, and implement the magnetic-ATS model in Section B, 
consider the 2D cartoon situation depicted in Fig. \ref{cellsT1T2}(a) for some temperature
$T_1 < T_g$: the intrinsic 
magnetisation is roughly proportional to the overall white-regions extension (per unit
area or mass). In fact it is proportional to the overall white-regions inner perimeter
length in 2D. Ignoring any slow-growth (or shrinking) of the oval RER size, let us change
quasi-statically the temperature to a nearby value $T_2$, giving rise to the slightly 
different equilibrium configuration of Fig. \ref{cellsT1T2}(b). It seems natural to expect 
that the $M_{intr}^{tunn}(T_2)$ will be only slightly different from (but not the same as) 
$M_{intr}^{tunn}(T_1)$ and that that moving slightly in temperature the oval RER will 
rotate slightly giving rise to oscillations as a function of $T$. More details and calculations 
in coming publications, but heuristically this mechanism explains the $M_{intr}(T)$ 
oscillations for fixed magnetic field $H$. Clearly, for a system of random-packed jammed 
spheres (or circles) all of the same size we do not expect any change of $M_{intr}(T)$ 
for fixed $H$ with chenging temperature in this picture. Thus, the oscillations in the intrinsic
magnetisation are evidence for the polydisperse size-distribution and non-spherical shape of 
the RER.

Interestingly, there might be here a connection with the so-called "Boson-peak" 
phenomenology \cite{SXL2021} as indeed the oscillations are at relatively high 
temperatures where the peak of the specific heat and of the vibrational spectrum 
$g(\omega)/\omega^2$ ($g(\omega)$ being the density of states at frequency $\omega$)
is observed. Then, as envisaged by one of us \cite{Jug2019}, the Boson-peak should
arise from the vibration modes of the closed-packed, jammed non-spherical RER 
ensemble.

Incidentally, the dynamics of the single RER may be the correct explanation also for 
recent findings near $T_g$ in a synchrotron-radiation XPCS experiment conducted on a 
sodium-silicate glass sample \cite{Ruta2014}. In such study below the nominal $T_g$  
faster than expected dynamics was observed with relaxation times tipically in the 
100 s range. An explanation might be offered by rotational diffusion of the compact
but non-spherical RER. More in forthcoming publications.

\begin{figure}[!h]
  \centering
  \vskip -0mm
  \includegraphics[scale=0.30] {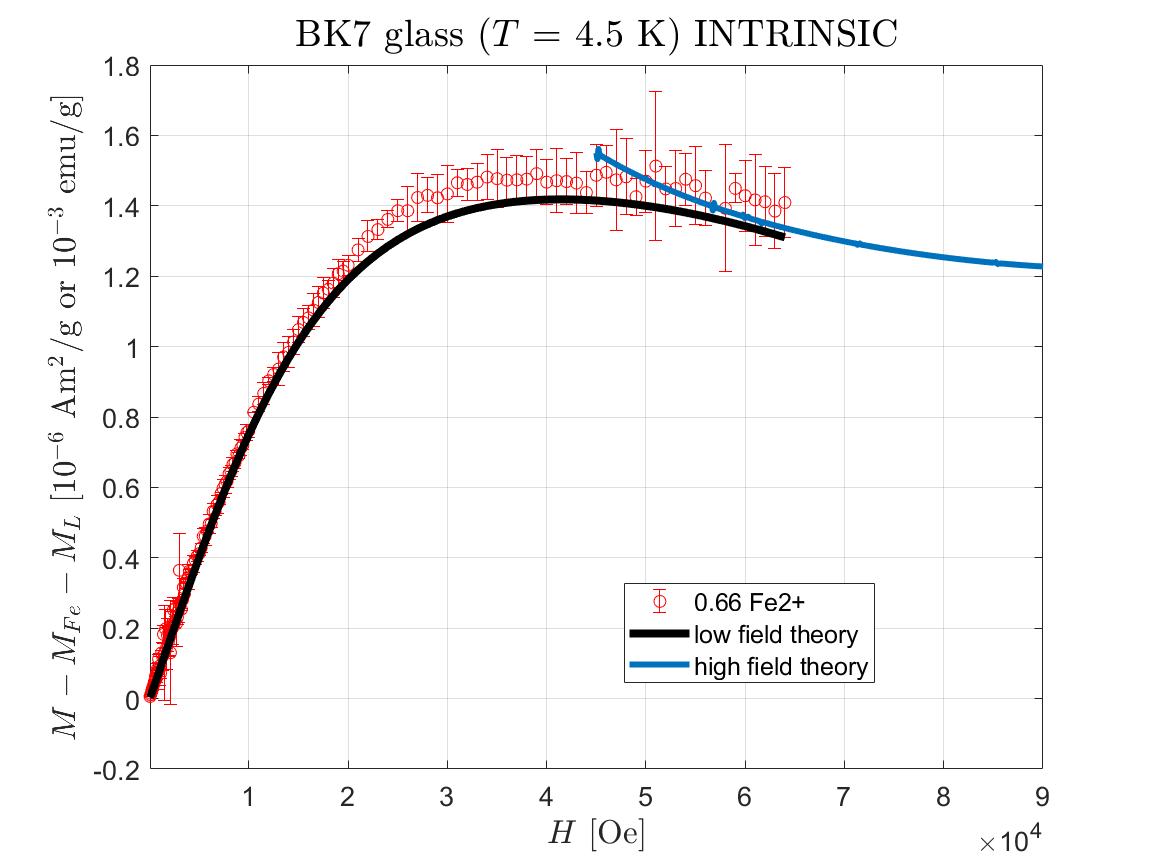}
 \vskip -0mm
\caption{Data for the subtracted (of $M_{LL}$ as in Eq. (1) PA) intrinsic magnetisation
for BK7 glass at 4.5 K. The data are reasonably well explained by our ETM theory, using 
the low- and intermediate-field approximation and the high-field approximation. The 
combined approximations give a conclusive descent for high values of $H$ after a 
Curie-like linear increase and broad peak due to the non-linear magnetic spectrum (Fig.
7(c), left panel, in the PA). The phenomenon of intrinsic glass paramagnetism is thus
completely new. 
}
 \label{Marcon}
\end{figure}

Another important challenge in explaining the data with the present theory is the
high-field values region. We fitted the data with the ETM theory discussed in detail in 
Section B, which strictly-speaking considers only two of the (at least) three energy levels 
of each single magnetic-sensitive ATS. This description is acceptable for intermediate 
$H$ values, but by following the descent of the effective ETM-theory $M_{intr}^{tunn}$ 
values one would arrive at the situation where a {\it negative} (thus diamagnetic) 
magnetisation occurs. This is indeed not realistic, and an improved high-$H$ theory
(taking all three levels of Hamiltonian (\ref{3lstunneling}) into account) shows that at 
high magnetic field the $M_{intr}^{tunn}$ calculated magnetisation smooths away to a 
slow descent with increasing high values of $H$ \cite{Mar2021}. In Fig. \ref{Marcon} we
present the situation for the BK7 glass at $T$=4.5 K, where the raw data are for the 
magnetisation after subtraction of the appropriate (MS determined Fe-concentration and 
best fits) Larmor and Langevin contributions. Both the low- and intermediate-magnetic 
field predicted $M_{intr}^{tunn}$ and the high-field $M_{intr}^{tunn}$ curves are drawn
from the present theoretical treatment and the match is satisfactory, certifying that the 
intrinsic magnetisation is expected to fall off gently with increasing high fields.

%----------------------------------------------------------------------------------------
%	SECTION 6
%----------------------------------------------------------------------------------------

\vskip 1.0cm

{\bf E. Novel Quantum-Coherence Phenomena.}

We finally comment on the very low-$T$ behaviour of $M_{intr}$. Very few data points
for $M_{intr}$ are available at such temperatures, but for BAS glass Fig. 5(a) (in the PA) 
already indicates interesting deviations below 4 K from the theoretical curve here obtained.
While for BK7 glass, the last experimental data point in Fig. 3 (in the PA) at the lowest
available temperature of 2 K might be off the present theory's curve because something 
new takes place for $T<$ 3 K. What could the new physics be?

Elaborating further on what stated in the PA, in reality evidence from the last 10 years or
so of study of the ETM applied to low-temperature data in mixed glasses suggests that 
the TLS in glasses sit at the interface between the solid-like RER and the fluid-like 
particles contained in their random-packing's ``voids'' \cite{Jug2013,Jug2010}. A better 
hypothesis is that, in fact, the TLS hide at the RER-RER interfaces and are precisely the 
degrees of freedom keeping the polycluster or cellular structure together when the 
same-polarisation charged chemical species in the ``voids'' would tend to make the 
polycluster structure fly apart. Within this scenario, then glasses are then truly frustrated 
systems (in a similar, but different sense as spin-glasses are \cite{FiHe1991}).

\begin{figure}[!h]
  \centering
  \vskip +10mm
  \includegraphics[scale=0.45] {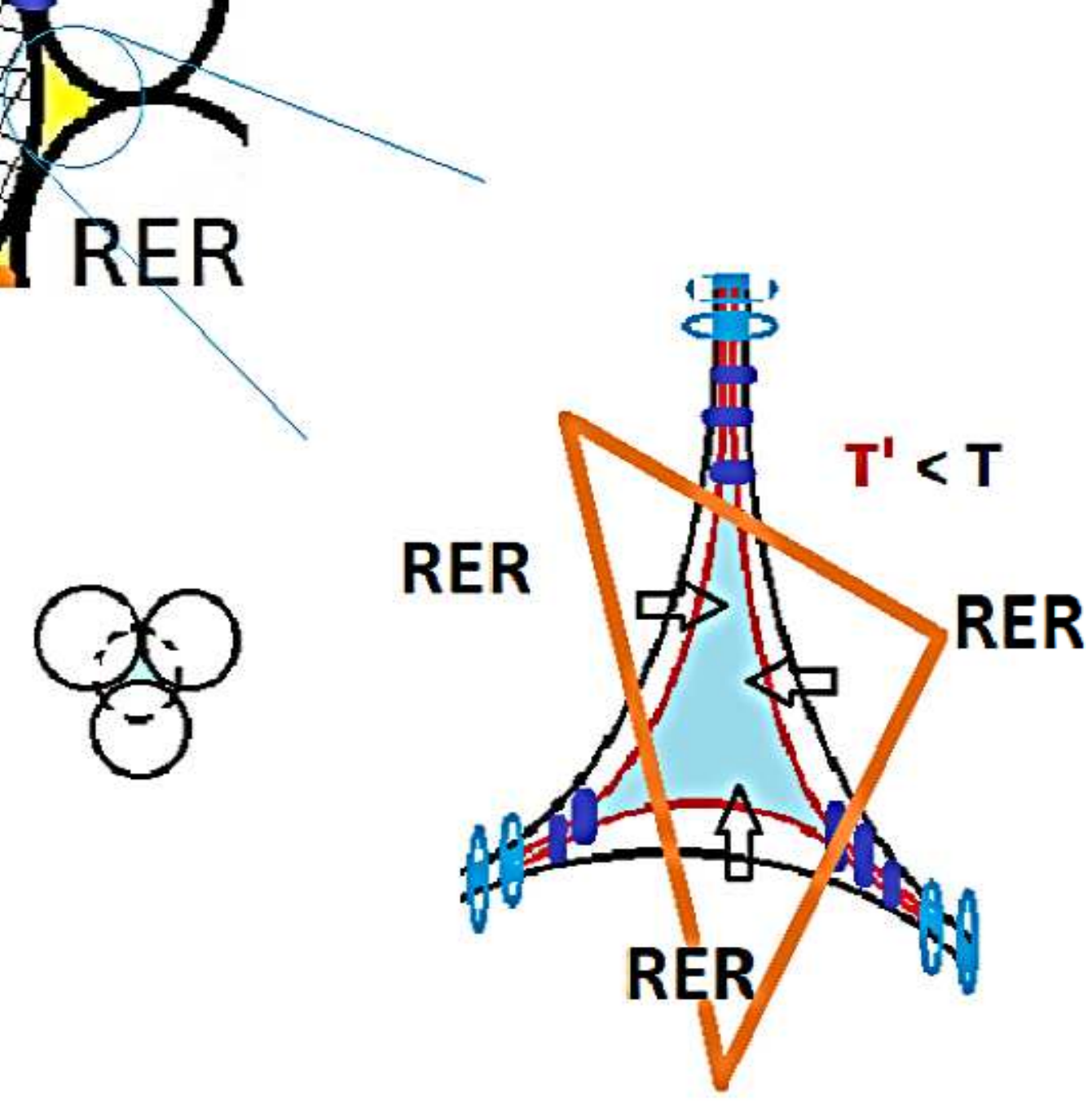}
 \vskip -20mm
\caption{ \big[What may really happen inside a RER-void (3D picture seen from above
as in the schematic packing of spheres)\big].
The microscopic origin of the ATS and of the TLS in bulk glasses: growth of the 
RER-RER interface through consolidation at the expense of the RER-void's mobile
particles (from black- to red-lines). As temperature lowers from $T$ to $T' < T$, 
the RER grow (arrows) into the ``void'' through adsorption of void's particles, so 
that the surface available to O$^{-}$ dangling bonds diminishes (light-blue area) 
while some more of the oxygens (dark-blue elements) end up as TLS in the 
RER-RER interface, which grows (zipping-up mechanism for the RER-RER interface). 
The orange triangle is schematic for the ATS quasiparticle (elementary excitation of
the light-blue O$^{-}$ region) three-welled potential. 
}
 \label{zippingup}
\end{figure}

However, as is argued in the caption of Fig. \ref{zippingup}, the TLS might indeed be the 
forthcomers of the O$^{-}$ dangling bonds sitting at the RER-void fuzzy interface which 
as temperature decreases gets to shrink further and further (see Section B). In fact, the 
TLS concentration increases in glasses at the lowest temperatures \cite{SXL2021}. Then, 
as temperature decreases the  O$^{-}$ ions get closer and closer together and the whole 
of the fluid contained in each void might become so strongly correlated that 3D coherent 
tunneling takes place. Therefore, the turning-up of the intrinsic magnetisation at the 
lowest temperatures for BAS-glass (and maybe also for BK7-glass) signals a 2D-3D
dimensional local-to-global crossover. At even lower temperatures, the tantalising 
possibility that the whole network of O$^{-}$ tunneling-ions-filled ``voids'' between the 
RER might become coherent would represent an extraordinary realisation of a natural 
quantum computer core containing some O(10$^{17}$) g$^{-1}$ qubits (order of 
magnitude of $n_{ATS}$). More investigations are under way.

%----------------------------------------------------------------------------------------
%	SECTION 7
%----------------------------------------------------------------------------------------

\vfill
{\bf G. Final Conclusions, Authors and Acknowledgements.}

As argued at length in this work, the phenomenon of {\it glass paramagnetism} is a 
reality and an unexpected new finding in the physics of glass with deep-reaching 
consequences for glass science in general. It is lamentable that it has been discovered 
in systems, the multi-silicates, where the level of Fe-impurity doping is always quite
high so that Langevin paramagnetism has to be subtracted away. Better would be to
conduct a systematic SQUID-magnetometry study in iron-group-free glass-forming 
systems like glycerol (C$_3$H$_8$O$_3$), with a $T_g$ of around 190 K. However: 
1) inserting the substance at the liquid state in the sample-holder is problematic (the 
container vessel is also made of a type of glassy material); 2) the resulting magnetisation 
is likely to be very very weak, because of the large size of the RER in the organic 
glasses \cite{Pal2011}; 3) there is very little knowledge about the values of the ETM 
parameters for glassy glycerol, none for other systems. We have thus chosen the 
multi-silicates because there has been extensive study of their unusual magnetic effects 
at low temperatures. 
Notice that ultra-pure silica glass (amorphous SiO$_2$) is on the other hand likely 
to be characterised by very small-sized RER because it is a mono-component 
glass-forming substance with O(1) nm-size dynamical heterogeneities \cite{JLT2021}.
Therefore, the O$^-$ dangling bonds on the RER surfaces will be strongly-correlated 
but too few in numbers ($N_{tunn}\sim$ O(10)), contrary to the case of the 
multi-silicate glasses. Consequently, as experimentally observed \cite{Lud2003}, there 
are no relevant magnetic-tunneling effects to be expected for {\it pure a}-SiO$_2$: they 
would be far too weak to be measurable because $N_{tunn}$ enters to the power 3 in
the combination 
$D_{0min,max}S_{\triangle}|Q/e|\propto N_{tunn}^3\Delta_{0min}a_0^2$ 
(Section B, recall that for the multisilicates $N_{tunn}\sim$ O(100) instead).

\vskip 0.5 cm
\underline{Authors:} this part of the work has been written up entirely by the PI (GJ) who 
is the owner of the ETM-theory, of the calculations and carried out the data
analysis and interpretation. As in the case of the PA, the qualitative MS chemical analysis 
was carried out by SR. 
\vskip 0.5 cm
\underline{Acknowledgements:} the PI is very grateful to Giacomo Marcon for help in 
working out the high-field behaviour of the intrinsic magnetisation. Also, the PI is grateful 
to Silvia Bonfanti for technical help and especially to Nika Fran{\v c}e{\v s}kin for 
encouragement in the completion of this work. Support by the INFN-Sezione di Pavia is
also gratefully acknowledged.

\vfill

%----------------------------------------------------------------------------------------
%	BIBLIOGRAPHY
%----------------------------------------------------------------------------------------

\newpage

\bibliographystyle{unsrt}

%----------------------------------------------------------------------------------------

\vfill
\end{document}